\documentclass[aip,jap,amsmath,amssymb,amsfonts,reprint]{revtex4-2}
\usepackage{graphicx}
\usepackage{hyperref}

\begin{document}

\title{Random 2D nanowire networks: Finite-size effect and the effect of busbar/nanowire contact resistance on their electrical conductivity } %Title of paper

\author{Yuri~Yu.~Tarasevich}
\email[Corresponding author: ]{tarasevich@asu.edu.ru}
\affiliation{Laboratory of Mathematical Modeling, Astrakhan State University, Astrakhan 414056, Russia}

\author{Andrei~V.~Eserkepov}
\email{dantealigjery49@gmail.com}
\affiliation{Laboratory of Mathematical Modeling, Astrakhan State University, Astrakhan 414056, Russia}

\author{Irina V. Vodolazskaya}
\email{vodolazskaya_agu@mail.ru}
\affiliation{Laboratory of Mathematical Modeling, Astrakhan State University, Astrakhan 414056, Russia}

\date{\today}

\begin{abstract}
We have studied the resistance of two-dimensional random percolating networks of zero-width metallic nanowires (rings or sticks). We toke into account the nanowire resistance per unit length, the junction (nanowire/nanowire contact) resistance, and the busbar/nanowire contact resistance. Using a mean-field approximation (MFA), we derived the total resistance of the nanoring-based networks as a function of their geometrical and physical parameters. We have proposed a way of accounting for the contribution of the busbar/nanowire contact resistance toward the network resistance. The MFA predictions have been confirmed by our Monte Carlo (MC) numerical simulations. Our study evidenced that the busbar/nanowire contact resistance has a significant effect on the electrical conductivity when the junction resistance dominates over wire resistance.
\end{abstract}

\maketitle %\maketitle must follow title, authors, abstract


\section{Introduction}\label{sec:intro}
In recent years, the study of random two-dimensional (2D) networks of conductive nanowires has become a hot topic of research. Due to the excellent electrical and optical performances of such systems, numerous applications are possible, while their low cost and ease of manufacturing promise very attractive prospects.\cite{Liu2019}

A comprehensive review of the different approaches used to describe the behavior of the electrical conductivity of random 2D nanowire networks can be found in Ref.~\onlinecite{Benda2019}. This review should be extended by some recently published works devoted to nanostick- and nanowire-based random 2D networks.\cite{Ponzoni2019APL,Kim2020,Fata2020JAP,Manning2020,Tarasevich2021JAP,Tarasevich2022PCCP,Tarasevich2022PRE}
\citet{Benda2019} proposed a closed-form approximation of the functional dependence of the effective resistance ($R_0$) of 2D random networks of zero-width stick nanowires on their physical parameters
\begin{equation}\label{eq:Benda}
R_0 = A R_\text{w} + B R_\text{j} + C R_\text{b},
\end{equation}
where $R_\text{w}$ is the wire resistance, $R_\text{j}$ is the junction resistance (wire-to-wire contact resistance), and $R_\text{b}$ is the busbar/wire contact resistance, while $A$, $B$, and $C$ are the geometrical coefficients. Similar relations have been proposed by other authors.\cite{Zezelj2012PRB,Kumar2017JAP,Ainsworth2018,Ponzoni2019APL} However, in those works, only the wire resistance and the junction resistance were taken into account, while the busbar/wire contact resistance was ignored. A consideration of the busbar/wire contact resistance as an independent and significant parameter is an important novelty of Ref.~\onlinecite{Benda2019}. This dependence has been derived using a dimensional analysis ($\Pi$-theorem). However, the coefficients $A$, $B$, and $C$ have been obtained using a fit of MC simulations rather than from an analytical consideration. These coefficients depend on both the geometry (the aspect ratio of electrode separation over stick length) and the number of wires with respect to the number of wires at the percolation threshold. However, the percolation threshold was not explicitly used in the derivation of Eq.~\eqref{eq:Benda}. By contrast, the percolation threshold and the two critical exponents were taken into account in Ref.~\onlinecite{Zezelj2012PRB}, viz., the sheet resistance of dense networks of randomly placed zero-width sticks is
\begin{equation}\label{eq:RsheetZS}
  R_\Box = \frac{b n^{t-1} R_\text{w} + (n + n_\text{c})^{t-2}R_\text{j}}{a\left[(n-n_\text{c})^t + c(L/l)^{-t/\nu}\right]},
\end{equation}
where $L$ is the linear system size, $l$ is the length of the stick, $n$ is the number density of the conductive sticks (number of sticks per unit area), $n_\text{c}$ is the percolation threshold, $a$, $b$, and $c$ are adjustable parameters, $\nu$ is the correlation-length exponent, and $t$ is the conductivity exponent.\cite{Zezelj2012PRB} Since the critical exponents as well the number of particles at the percolation threshold are dimensionless, these quantities can readily be incorporated into dimensional analyse.

The resistance of the contacts of the conductive wires with the busbars was taken into account in computations by \citet{Forro2018ACSN} However, their conductivity estimates using an MFA ignored this resistance. To reconcile the MFA predictions and computations, the authors introduced a correction factor (a so-called ``effective wire length''). In our own recent computations, the busbar/wire resistance was 0, so the consistency of the computations and the MFA predictions was provided without any correction factor being required.\cite{Tarasevich2022PRE}

In the case of nanoring-based films, when the junction resistance dominates over the wire resistance while the busbar/nanowire contact resistance and the junction resistance are of the same order of magnitude, the MFA significantly overestimates the electrical conductivity when compared to the computations.\cite{Tarasevich2021JAP} Thus, the contribution of the busbar/wire resistance to the sheet resistance seems to be important.

The increased effective length associated with microstructured electrodes results in modifications of the electrical device behavior within the same Ag nanowire network.\cite{Fairfield2014} Reduction of 10--40\% was observed in the sheet resistance, with the strongest reductions for devices with two serrated electrodes. The nanowires had mean diameters of $85 \pm 10$~nm and lengths of $7.5 \pm 2\,\mu$m. Active network areas were $50 \times 50$ and $200 \times 200\,\mu$m. Hence, the values of the ratio $L/l$ were from 6.67 to 26.67. For the flat electrode devices, the authors reported $R = 4800$~$\Omega$. Note that the sizes of nanowire-based films can be very different, viz., $18 \times 18$~mm,\cite{Kou2017} $10 \times 30$~mm,\cite{Park2016} $120 \times 20\,\mu$m.\cite{Rocha2015}

Reported data evidenced that the ratio of Ag nanowire length to its diameter ranges from 100 to 1000 of the order of magnitude (Table~\ref{tab:NR}), i.e., a slender-rod approximation seems to be valid. Four-probe measurements on almost 40 individual nanowires with diameters ranging from 50 to 90~nm gave an average resistivity of $ 20.3\pm 0.5$~n$\Omega\cdot$m.\cite{Bellew2015} \citet{Rocha2015} reported similar value of the average Ag nanowire resistivity, viz.,  $22.6 \pm 2.3$~n$\Omega\cdot$m obtain using 15 Ag nanowires.\cite{Rocha2015} Using three different methods, the junction resistance for Ag nanowires was determined. The distribution shows a strong peak at $11\,\Omega$, corresponding to the median value of the distribution.\cite{Bellew2015} \citet{Selzer2016} reported that a value of the resistance of a single Ag nanowire is $4.96 \pm 0.18\,\Omega/\mu$m while the junction resistance is $25.2 \pm 1.9\,\Omega$ (annealed junctions) and $529 \pm 239\,\Omega$ (non-annealed ones). Out estimates (the last column of Table~\ref{tab:NR}) evidenced that, typically, the junction resistance is significantly larger than the wire resistance (up to 2 orders of magnitude), $R_\text{j} \gg R_\text{w}$.
\begin{table*}
  \caption{Published experimental data on Ag nanowires along with our estimates (last two columns) of the aspect ratio and the wire resistance based on these data.}\label{tab:NR}
  \centering
  \begin{ruledtabular}
  \begin{tabular}{lcccc}
    Reference  & length $(l)$, $\mu$m & diameter $(d)$, nm & $l/d$ & $R_\text{w}$, $\Omega$ \\
    &&&& ($ \rho_\text{w} = 20.3$~n$\Omega\cdot$m)\\
           \hline
     \citet{Lee2008} & $ 8{,}7\pm 3{,}7$ & $ 103\pm 17$ & $\approx 84$ & $\approx 21$ \\
     \citet{Nguyen2019} & $ 7\pm 3$ & $ 79\pm 10$ & $\approx 88$ & $\approx 29$ \\
    \citet{Khanarian2013} & 4{,}8--36 & 56--153 & 31--360 & $\approx 48$ \\
     \citet{He2018} & $66$ & $160$ & $\approx 410 $ & $\approx 67$ \\
     \citet{Rocha2015} & $6.7$ & $50 \pm 13$ & $134$ &  $\approx 69$ \\
      \citet{Selzer2016} & $25$ & $90$ & $278$ &  $\approx 80$ \\
     \citet{Bellew2015} & $ 7\pm 2$ & $ 42\pm 12$ & $\approx 167$ & $\approx 103$ \\
     \citet{Kou2017} & $\approx 15$ & $\approx 60$ & $\approx 250$ & $\approx 108$\\
     \citet{Xu2018} & $ 123$ & 120 & $\approx 1000 $ & $\approx 221$ \\
     \citet{Oh2018} & $22 \pm 5$ & $25 \pm 3$ & $\approx 880$ &  $\approx 910$ \\
     \citet{Lee2016} & $ 20$ & 16--22 & $\approx 1000$ & $\approx 1432$ \\
     \end{tabular}
   \end{ruledtabular}
\end{table*}

Silver nanorings with the uniform ring diameter of $15 \pm 5\,\mu$m can also be synthesized.\cite{Azani2018,Azani2019} Depending on the nanoring concentration, the sheet resistance of nanoring-based films of size $5 \times 5$~cm varies from 20 to 350 $\Omega/\Box$.\cite{Azani2019} In these samples, the ratio of the linear system size to the nanoring diameter is approximately 3000.

The goal of our study was to evaluate the contribution of the busbar/nanowire contact resistance in the total resistance of 2D systems of randomly placed conductive wires. For this purpose, we derived a closed-form expression for the effective electrical resistance as a function of the main physical parameters including the busbar/nanowire contact resistance. An MC simulation has been performed to test this expression.

The rest of the paper is constructed as follows. Section~\ref{subsec:simul} is devoted to our simulation. Since details of the computer simulation have been described previously,\cite{Azani2019,Tarasevich2021JAP,Tarasevich2022PRE,Tarasevich2022PCCP} we present in Sec.~\ref{subsec:simul} only a brief sketch using the nanoring-based system as an example. The electrical conductivity of the system of randomly placed conductive rings using a discreet version of the MFA is derived in Sec.~\ref{subsec:MFArings}. The effect of the busbars on the electrical resistance is analyzed in Sec.~\ref{subsec:busbars}. In Section~\ref{sec:results}, we compare our MC simulations and theoretical predictions. Section~\ref{sec:concl} summarizes the main results and discusses some open questions.

\section{Methods}\label{sec:methods}
\subsection{Computer simulation}\label{subsec:simul}
Sampling is very close to the used previously.\cite{Azani2019,Tarasevich2021JAP,Tarasevich2022PCCP,Tarasevich2022PRE}
Identical conductive fillers were randomly placed on a substrate. Their centers were independent and identically distributed within a square domain of size $L \times L$. To reduce the finite-size effect, periodic boundary conditions (PBCs) were applied along both mutually perpendicular directions. Since the electrical conductivity is our primary interest, the number density was required to be above the percolation threshold, $n \geqslant n_\text{c}$. When the desired number density of the fillers was reached, the PBCs were removed, allowing us to consider the model as an insulating film of size $ L \times L $ covered by conductive fillers. Two opposite borders of the domain were considered as superconductive busbars. A potential difference, $V_0$, was applied to these busbars. The electrical resistance per unit length of each filler is $\rho_\text{w}$. The electrical resistance of each contact (junction) between any two fillers is $R_\text{j}$. The electrical resistance of each contact (junction) between a filler and a busbar is $R_\text{b}$. Junctions are assumed to be ohmic. The system under consideration can be treated as a random resistor network (RRN). Such a network is an irregular network with different branch resistances. Applying Ohm's law to each branch and Kirchhoff's point rule to each junction, a system of linear equations (SLEs) can be written. Although this SLE is huge, its matrix is sparse. We used the EIGEN library\cite{eigenweb} to solve it numerically.

Two particular kinds of fillers were considered, viz, rings with a given radius ($r=1$,  $R_\text{w} = 2 \pi r \rho_\text{w}$) and equiprobable orientated zero-width sticks with a given length ($l=1$).

Generally, we used domains of the fixed size $L=32 $. An additional study of the finite-size effect was performed. To efficiently determine the percolation threshold (occurrence of a percolation cluster that spans the system in a given direction), the union-find algorithm~\cite{Newman2000PRL,Newman2001PRE} was used. In our simulations, $n_\text{c} = 0.373 \pm 0.004$ for rings, while $n_\text{c} = 5.641 \pm 0.025$ for sticks.

Computations for systems of randomly placed zero-width conductive sticks were carried out similarly. Depending on which effect was of interest, in the figures we used either the resistance or the electrical conductivity as alternatives. The results of the computations were averaged over 100 independent runs. The error bars in the figures correspond to the standard deviation of the mean. When not shown explicitly, they are of the order of the marker size.

\subsection{Mean-field approach for randomly placed conductive rings}\label{subsec:MFArings}
\begin{figure}[!htb]
  \centering
  \includegraphics[width=\columnwidth]{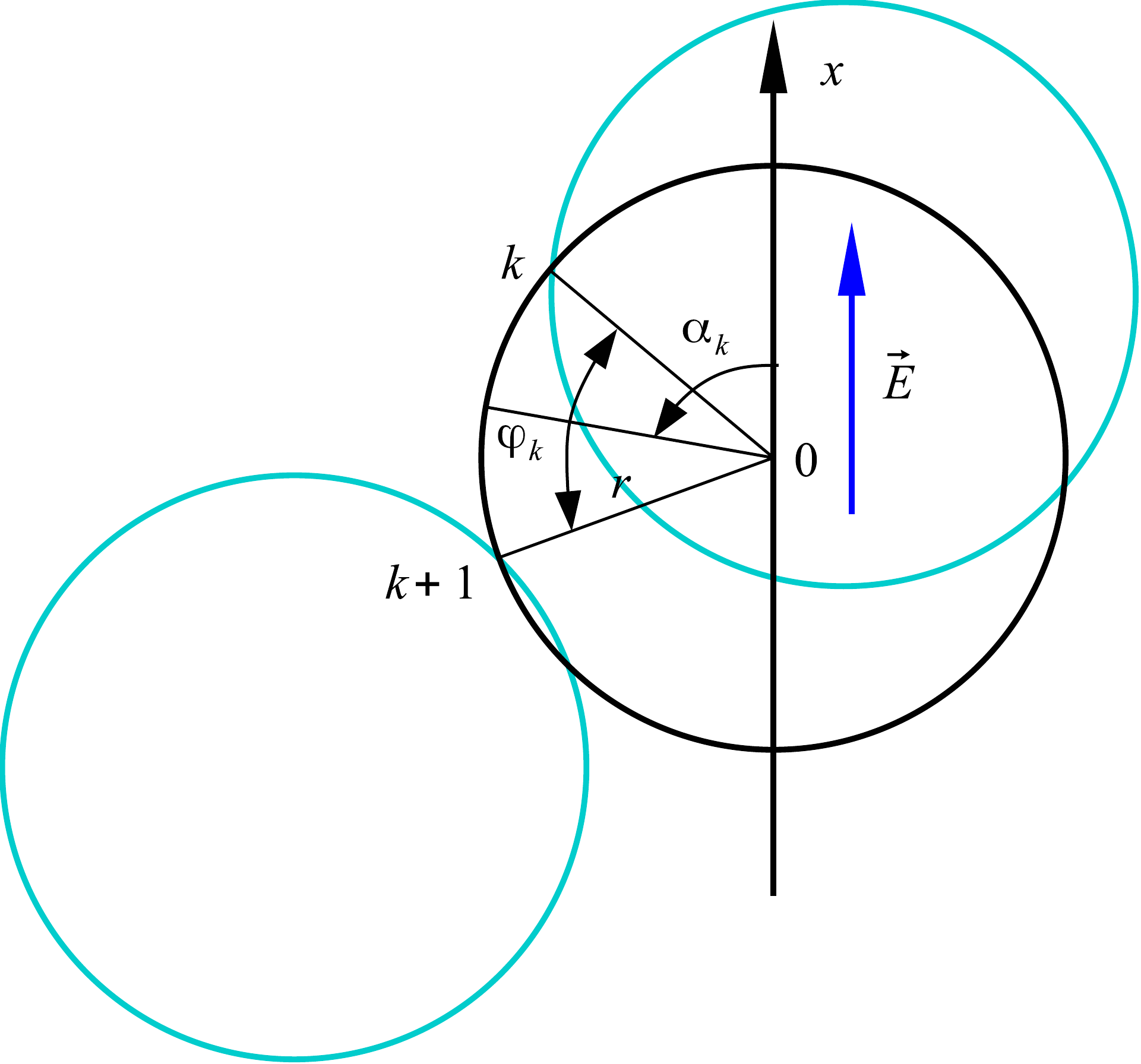}
  \caption{Conductive ring in an electric field. See explanations in the text.}\label{fig:ring}
\end{figure}

When the number density of the conductive rings is large enough, the variation of the electrical potential along the film is close to linear.\cite{Azani2019} The basic idea behind the MFA is as follows. Only one conductive ring in the mean-field produced by all the other rings can be considered instead of a consideration of the entire network. Let there be a conductive ring of radius $r$ placed in an external electrostatic field $\mathbf{E}$  (Fig.~\ref{fig:ring}). The potential of the external field depends on the coordinate, $V(x)$. The conductive ring is assumed to be covered with an ideal insulator. Each intersection of this ring with other rings (junctions) in the original system is considered in the MFA as an insulation imperfection that can lead to some leakage current. Thus, each junction is assumed to be possessed of an electrical resistance $R_\text{j}$.

Consider the arc of the ring between the $k$-th and the $k+1$-th contacts (Fig.~\ref{fig:ring}). Let the central angle of this arc be $\varphi_k$, while the angle between the direction of the external electrostatic field and the direction from the ring center to the center of the arc be equal to $\alpha_k$. The resistance of the arc $\varphi_k$ between these two contacts is equal to $\varphi_k r \rho_\text{w}$.

According to Ohm's law, the potential drop along the arc is
\begin{equation}\label{eq:Ohm}
u_{k+1} - u_k+ i_k \varphi_k r \rho_\text{w} = 0,
\end{equation}
where $ u_k $ and $u_{k+1}$ are the potentials of the contacts $k$ and $k+1$, respectively, while $i_k$ is the electrical current in the arc between these contacts. According to Kirchhoff's rule, for the $k$-th contact
\begin{equation}\label{eq:v}
i_k - i_{k-1} + \frac{u_k - V_k}{R_\text{j}} = 0,
\end{equation}
where $V_k$ is the potential of the external field at the point where the $k$-th contact is located. Hence, the potential of the $k$-th contact, $u_k$, can be written as
\begin{equation}\label{eq:uk}
 u_k = V_k - (i_k - i_{k-1})R_\text{j}.
\end{equation}
The potential of the $k+1$-th contact, $u_{k+1}$, can be written similarly.
Substitution of the potential drop between these two contacts into Eq.~\eqref{eq:Ohm} leads to
\begin{equation}\label{eq:Ohmnew}
V_{k+1} - V_k - (i_{k+1} - 2 i_{k} + i_{k-1})R_\text{j}+ i_k \varphi_k r \rho_\text{w} = 0.
\end{equation}
Taking the center of the ring as the zero potential of the external field, we get the dependence of the potential on the angular coordinate
\begin{equation}\label{eq:Vbeta}
V(\beta) = - E r\cos \beta,
\end{equation}
where $\beta$ is the angle between the direction of the electrostatic field and the radius vector that connects the center of the ring with the given point. Accordingly, the potential of the external field at the $k$-th contact is
\begin{equation}\label{eq:Vk}
V_k = - E r \cos \left(\alpha_k - \frac{\varphi_k}{2} \right),
\end{equation}
while the potential of the external field at the $k+1$-th contact is
\begin{equation}\label{eq:Vkplus1}
V_{k+1} = - E r \cos \left(\alpha_k + \frac{\varphi_k}{2} \right).
\end{equation}
Then,
\begin{equation}\label{eq:Vdiff}
V_{k+1} - V_k = 2 E r \sin \alpha_k \sin \frac{\varphi_k}{2}.
\end{equation}
Hence, Eq.~\eqref{eq:Ohmnew} can be rewritten as
\begin{multline}\label{eq:Ohmnew1}
2 E r \sin \alpha_k \sin \frac{\varphi_k}{2}\\ - (i_{k+1} - 2 i_{k} + i_{k-1})R_j+ i_k \varphi_k r \rho_\text{w} = 0.
\end{multline}
To solve this recurrence relation, some additional assumptions are required.

Let $R_\text{j}=0$. Then,
\begin{equation}\label{eq:reccJDA}
2 E \sin \alpha_k \sin \frac{\varphi_k}{2} + i_k \varphi_k \rho_\text{w} = 0.
\end{equation}
According to Ref.~\onlinecite{Akhunzhanov2020}, the probability density function (PDF) for the arc's central angle is
\begin{equation}\label{eq:PDF}
f(\varphi_k) = 4nr^2 \exp(-4nr^2\varphi_k)
\end{equation}
when $n r^2 \gg 1$. Multiply equation~\eqref{eq:reccJDA} by the PDF~\eqref{eq:PDF} and then integrate over all angles
\begin{equation}\label{eq:reccJDAint}
 \frac{16 n r^2 E }{64 n^2 r^4 +1} \sin \alpha_k + i_k \frac{\rho_\text{w}}{4 n r^2} = 0.
\end{equation}
Thus, the electrical current in the $k$-th arc is
\begin{equation}\label{eq:ikJDA}
i_k = -\frac{64 n^2 r^4 E }{(64 n^2 r^4 +1)\rho_\text{w} } \sin \alpha_k \approx -\frac{ E } {\rho_\text{w} } \sin \alpha_k .
\end{equation}
The average current at the point of the ring that has the angular coordinate $\beta$, is equal to
\begin{equation}\label{eq:meaniJDA}
\langle i(\beta) \rangle = 4 n r^2\int\limits_{\beta - \langle \varphi \rangle}^{\beta} i_k \, d \alpha_k \approx -\frac{ E }{\rho_\text{w} }\sin \beta
\end{equation}
or, in Cartesian coordinates,
\begin{equation}\label{eq:ixJDA}
  \langle i(x) \rangle \approx
  \frac{E }{\rho_\text{w} }\sqrt{1-\left(\frac{x}{r}\right)^2} .
  \end{equation}
The total electrical current that is carried by all the rings intersecting a
given equipotential line, is
\begin{equation}\label{eq:ItotalJDA}
I = 2 n L \int\limits_{-r}^{r} \langle i(x) \rangle \, dx \approx \frac{ \pi n r E L}{\rho_\text{w} }.
\end{equation}
According to the continuum (vector) form of Ohm's law $I = \sigma E L$. Hence, the formula for the electrical conductance is
\begin{equation}\label{eq:sigmaJDA}
\sigma = \frac{\pi n r}{ \rho_\text{w} }.
\end{equation}
This formula coincides with that which was obtained within the framework of the continuous approximation.\cite{Tarasevich2021JAP}

Now, let us assume that the contacts are evenly distributed over the ring. Hence, the angle between any two nearest contacts is $\left( 4nr^2\right)^{-1}$, while
$\sin \varphi_k \approx \left(4nr^2\right)^{-1}$ and $1 - \cos \varphi_k \approx (32 n^2 r^4)^{-1}$.
Thus, Eq.~\eqref{eq:Ohmnew1} transforms in
\begin{equation}\label{eq:Ohmaver}
\frac{ E}{4nr} \sin \alpha_k - (i_{k+1} - 2 i_{k} + i_{k-1})R_j+ i_k \frac{\rho_\text{w}}{4nr} = 0.
\end{equation}
We will look for a solution to the recurrence relation~\eqref{eq:Ohmaver} in the form
$i_k = A \sin \alpha_k$.
Substitution of this guest solution into Eq.~\eqref{eq:Ohmaver} leads to
\begin{equation}\label{eq:coeefs}
 A = -\frac{4 n r^3 E}{ R_\text{j} + 4 n r^3\rho_\text{w}},
\end{equation}
since $\sin\alpha_{k-1} = \cos \varphi_k \sin\alpha_{k}- \sin \varphi_k \cos\alpha_{k}$ and
$\sin\alpha_{k+1} = \cos \varphi_k \sin\alpha_{k} + \sin \varphi_k \cos\alpha_{k}$.
%=================================
Thus, the electrical current in the $k$-th arc is
\begin{equation}\label{eq:ik}
i_k = -\frac{4 n r^3 E}{ R_\text{j} + 4 n r^3 \rho_\text{w}} \sin \alpha_k .
\end{equation}
The total electrical current that is carried by all the rings intersecting a given equipotential line, is
\begin{equation}\label{eq:itntegral}
I = \frac{ 4\pi n^2 r^4 L E }{ R_\text{j} + 4 n r^3 \rho_\text{w}}.
\end{equation}
The electrical conductance is
\begin{equation}\label{eq:sigma}
  \sigma = \frac{4\pi n^2 r^4 }{ R_\text{j} + 4 n r^3\rho_\text{w}},
\end{equation}
while the electrical resistance is
\begin{equation}\label{eq:R}
  R = \frac{ R_\text{j}}{4\pi n^2 r^4 } + \frac{\rho_\text{w}}{\pi n r }.
\end{equation}
Formula~\eqref{eq:sigma} coincides with the formula obtained using a continuous approach.\cite{Tarasevich2021JAP}

\subsection{Effect of busbars}\label{subsec:busbars}
We will distinguish between the potential difference applied to the busbars, $V_0$, and the potential difference, $EL$, between the boundaries of the internal region of the system under consideration. Only when the resistance of the contacts of the conductive wires (sticks, rings, etc.) with the busbars is zero, does $V_0 = EL$. Otherwise, the voltage drops at the contacts of the conductive wires with the busbars should be taken into consideration. The average voltage drop at each busbar/wire contact is
\begin{equation}\label{eq:Deltau}
\Delta u = \frac{I R_\text{b}}{N_\text{b}},
\end{equation}
where $N_\text{b}$ is the number of contacts between the wires and the busbar.
Since there are two busbars, $EL + 2\Delta u = V_0$.
Accounting for the continuum form of Ohm's law,
\begin{equation}\label{eq:Ibarcommon}
I = \sigma \left( V_0 - \frac{2 I R_\text{b}}{N_\text{b}} \right).
\end{equation}
Hence, the effective resistance accounting for the busbar/wire contact resistance is
\begin{equation}\label{eq:Reffrings}
R_0 = \frac{ N_\text{b} + 2 \sigma R_\text{b}}{\sigma N_\text{b}}= R + \frac{2 R_\text{b}}{ N_\text{b}}.
\end{equation}
In our particular case, $N_\text{b} = 4 r n L$, hence,
\begin{equation}\label{eq:Reffrings1}
R_0 = \frac{2 r n L + \sigma R_\text{b}}{2 \sigma r n L}.
\end{equation}

Substituting~\eqref{eq:sigma} into~\eqref{eq:Reffrings1}, we get the formula
\begin{equation}\label{eq:Refflim}
R_0 = \frac{\rho_\text{w}}{\pi n r } + \frac{ R_\text{j}}{4\pi n^2 r^4 } + \frac{ R_\text{b}}{2 r n L}.
\end{equation}
Both the system size and the busbar/nanowire resistance are presented only in the last term. Hence, the last term describes both a finite-size effect and the effect of the busbar/nanowire resistance on the electrical conductivity. This term is especially significant when $ R_\text{j} \gg R_\text{w}$ [the so-called ``junction dominated approach'' (JDA)\cite{Manning2019} or, alternatively, the ``junction resistance dominant assumption'' (JRDA)\cite{Kim2020}]. Comparison of~\eqref{eq:Benda} and~\eqref{eq:Refflim} clearly evidenced that $A \sim n^{-1}$, $B \sim n^{-2}$, and $C \sim n^{-1}$. In the thermodynamic limit ($L \to \infty$), the resistance of the system is determined only by the two first terms of formula~\eqref{eq:Refflim}, i.e., it is independent of the resistance of the contacts of the conductors with the busbars.

Thus, for a 2D nanoring-based random system, an analog of Eq.~\eqref{eq:Benda} is obtained. In contrast with Ref.~\onlinecite{Benda2019}, the explicit expressions for the geometrical coefficients are presented for the system under consideration.

The above method can easily be transferred to other 2D systems, e.g., random metallic
nanowire networks with zero-width rodlike wires (sticks).\cite{Tarasevich2022PRE,Tarasevich2022PCCP} For a dense system,\cite{Tarasevich2022PRE,Tarasevich2022PCCP}
\begin{equation}\label{eq:MFAsigma}
\sigma = \frac{ \langle N_\text{j} \rangle }{2 C R_\text{w}} \left[ 1 - \sqrt{\frac{4 }{ \langle N_\text{j} \rangle \Delta} } \tanh\left(\sqrt{\frac{ \langle N_\text{j} \rangle \Delta}{4} }\right)\right],
\end{equation}
where
\begin{equation}\label{eq:Delta}
\Delta = \frac{R_\text{w}}{R_\text{j}}, \quad \left\langle N_\text{j}\right\rangle = n l^2 C, \quad C = \frac{2}{\pi}.
\end{equation}
In this case, the number of contacts between the wires and a busbar is \begin{equation}\label{eq:Nbstick}
N_\text{b} = C n l L,
\end{equation}
 where $l$ is the stick length. Hence,
\begin{equation}\label{eq:Reffstick}
  R_0 = \frac{ 1}{\sigma } + \frac{\pi R_\text{b}}{ n l L}.
\end{equation}
Using a series expansion of $R = \sigma^{-1}$~\eqref{eq:MFAsigma}, the effective resistance can be split into three terms like~\eqref{eq:Benda}
\begin{equation}\label{eq:split}
   R_0 = \frac {12  \rho_\text{w} }{5 n l} +\frac {12 \pi   R_\text{j} }{ n^2 l^4} + \frac{\pi R_\text{b}}{ n l L}.
\end{equation}
Hence, the finite-size effect and the effect of the busbar/nanowire resistance on the electrical conductivity can be studied independently of other effects. Again, comparison of~\eqref{eq:Benda} and~\eqref{eq:split} evidenced that $A \sim n^{-1}$, $B \sim n^{-2}$, and $C \sim n^{-1}$. In the thermodynamic limit ($L \to \infty$), the resistance of the system is determined only by the two first terms of formula~\eqref{eq:split}, i.e., it does not depend on the resistance of the contacts of the conductors with the busbars. When the junction resistance dominates over the wire resistance, Eq.~\eqref{eq:split} is simplified, then
\begin{equation}\label{eq:RstickJDR}
  R_0 = \frac{ 12 \pi R_\text{j}}{n^2 l^4 } + \frac{\pi R_\text{b}}{ n l L}.
\end{equation}

\section{Results}\label{sec:results}

\subsection{Ring-based conductive films}\label{subsec:RBCF}

\begin{figure}[!htbp]
  \centering
  \includegraphics[width=\columnwidth]{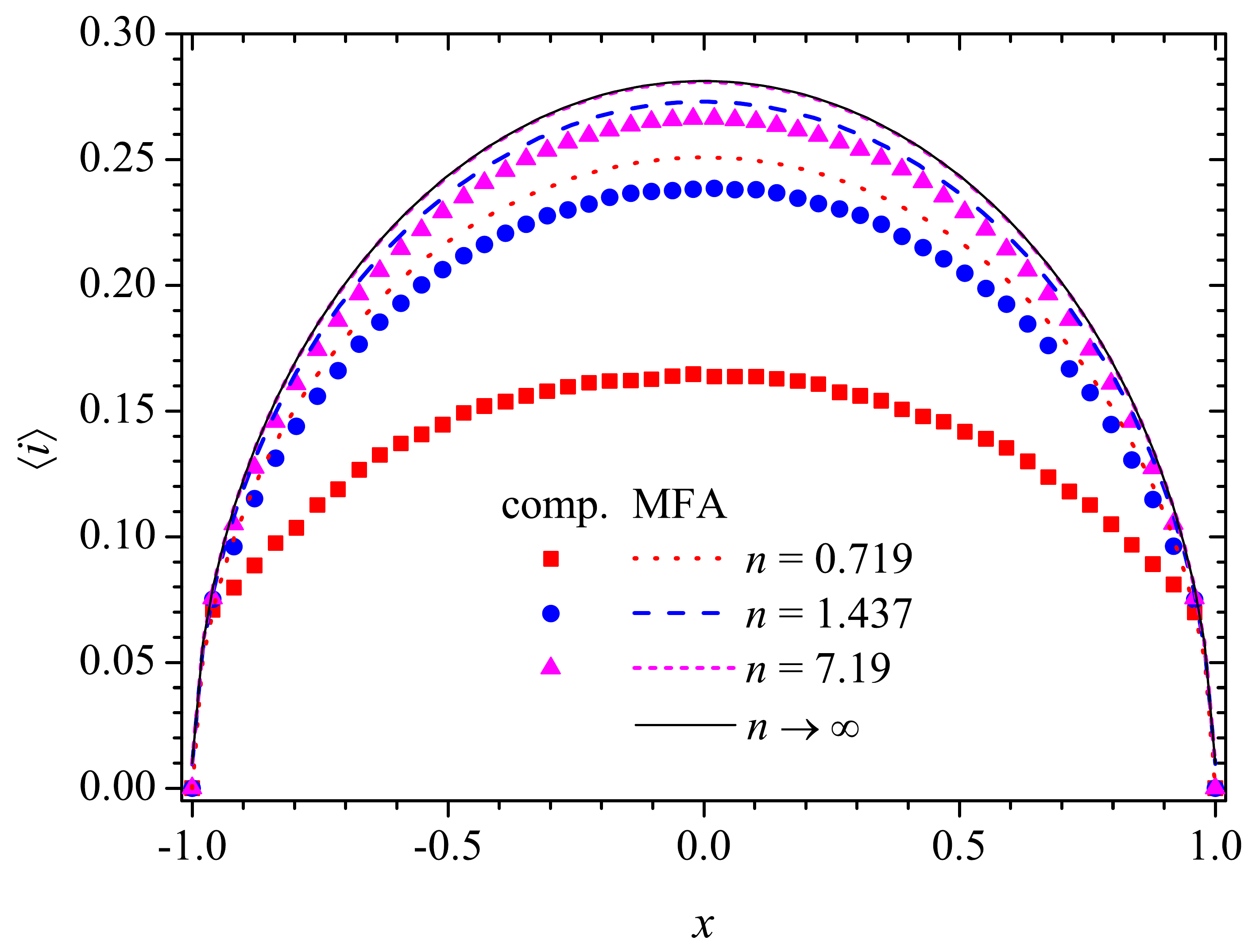}
  \caption{Dependencies of the electrical current, $\langle i \rangle$, on the position in the conductive ring, $x$, for different values of the number density. Here 0 corresponds to the ring center. The current is averaged over all the rings in one sample. The lines correspond to the MFA prediction~\eqref{eq:ixJDA}.}\label{fig:currentsWDRrings}
\end{figure}

Figure~\ref{fig:currentsWDRrings} presents the dependencies of the average electrical current, $\langle i(x) \rangle$, on the position in the conductive ring, for different values of the number density. Here, 0 corresponds to the ring center. The lines correspond to the MFA prediction~\eqref{eq:ixJDA}. Figure~\ref{fig:currentsWDRrings} evidenced that the MFA prediction is reasonable only for sufficiently dense systems. This is quite expected, since the assumption $n r^2 \gg 1$ was used in the derivation.

\begin{figure}[!htbp]
  \centering
  \includegraphics[width=\columnwidth]{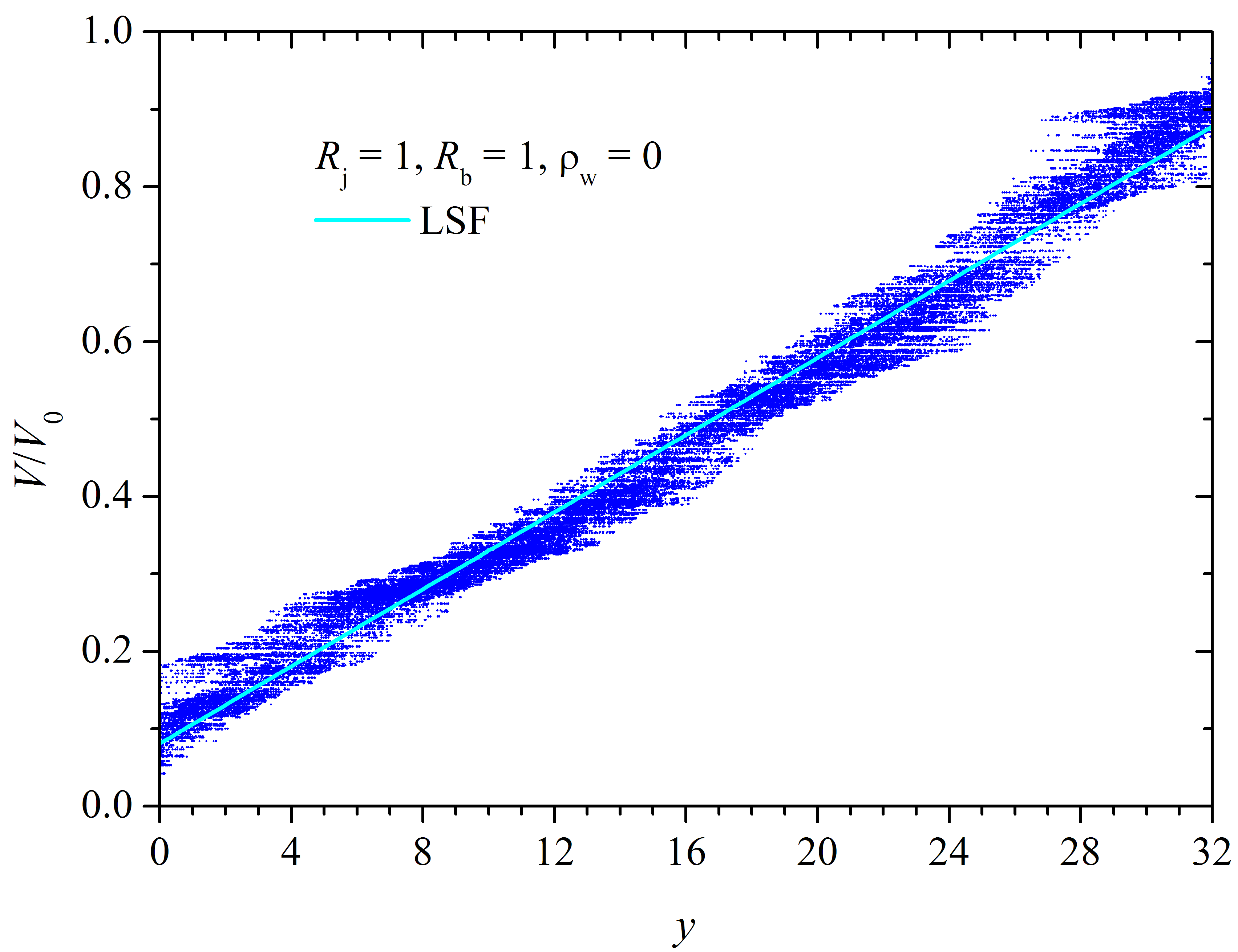}
  \caption{Example of potential distribution in one particular sample with randomly placed conductive rings. The potential of each junction in the system is plotted here against its position in the sample, $y$. $R_\text{j} = 1$, $R_\text{b} = 1$, $\rho_\text{w} = 0$, $n = 1.455$.}\label{fig:potential}
\end{figure}

Figure~\ref{fig:potential} shows the potential distribution in one particular sample with randomly placed conductive rings ($R_\text{j} = 1$, $R_\text{b} = 1$, $\rho_\text{w} = 0$). The potential drop along the sample is close to linear. The potential of each junction in the system is plotted here against its position in the sample. However, although the busbar potentials are 0 and 1, the potentials of the contacts closest to the busbars differ markedly from these values.

\begin{figure}[!htbp]
  \centering
  \includegraphics[width=\columnwidth]{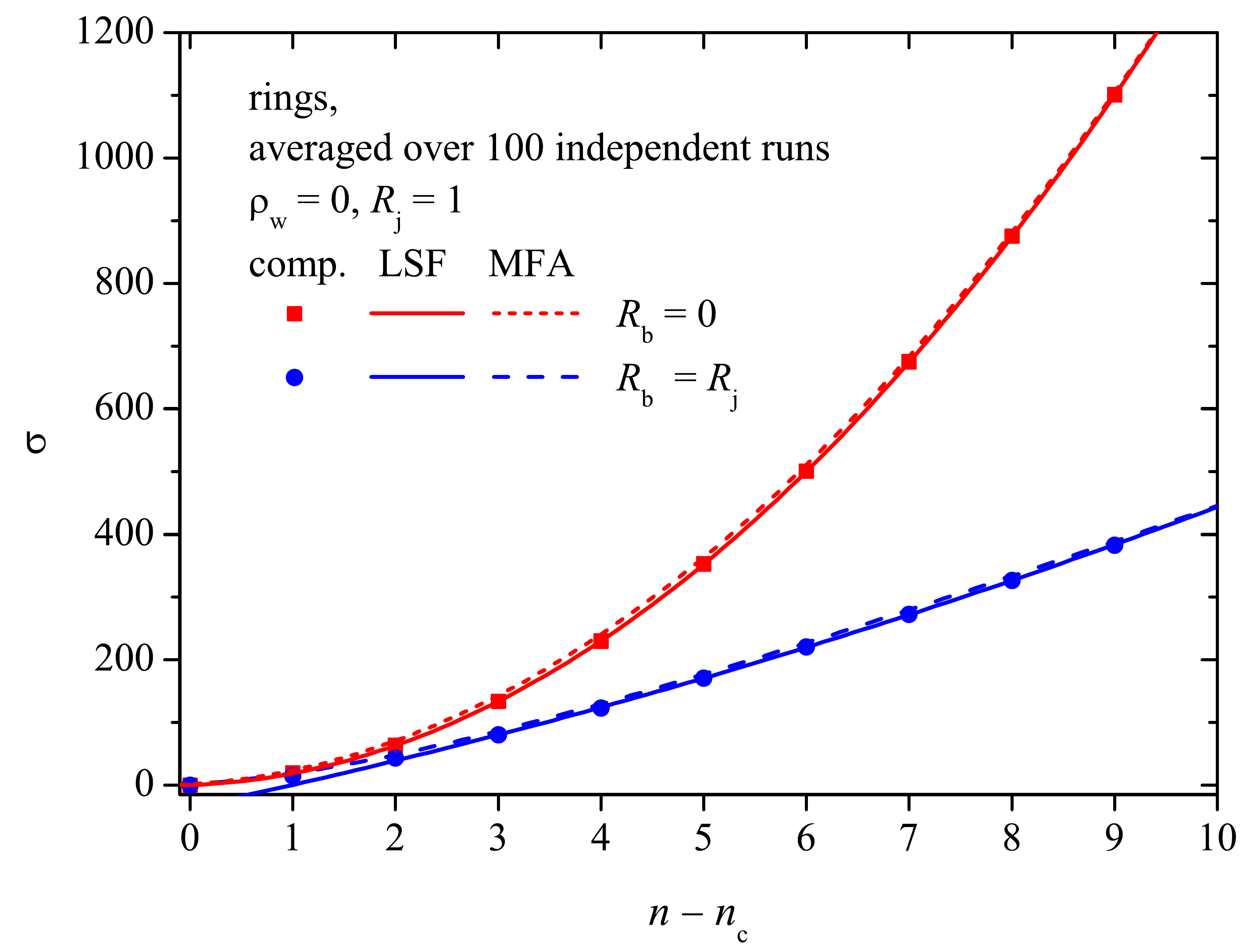}
  \caption{Dependence of the electrical conductance on the number density of the conductive rings, $\rho_\text{w}=0$, $R_\text{j} =1$. The resistances between a conductive ring and a busbar are $R_\text{b} = 0$ (squares) and $R_\text{b} = 1$ (circles). The markers correspond to the direct computation of the electrical conductance. The solid lines correspond to the least squares approximation by a polynomial of the second degree. The dashed lines correspond to the mean-field approximation (Eq.~\eqref{eq:RefflimJDRbus}).}\label{fig:JDR100}
\end{figure}

Figure~\ref{fig:JDR100} exhibits the electrical conductance against the number density of the conductive rings when the junction resistance dominates over the wire resistance ($\rho_\text{w}=0$, $R_\text{j} =1$). In this particular case, Eq.~\eqref{eq:Refflim} leads to the electrical conductance
\begin{equation}\label{eq:RefflimJDRbus}
\sigma_0 = \frac{4\pi n^2 r^4 L }{ R_\text{j}L + 2 \pi n r^3 R_\text{b}}.
\end{equation}
The markers correspond to the direct computation of the electrical conductance, viz., $R_\text{b} = 0$ (squares) and $R_\text{b} = 1$ (circles) (see Section~\ref{subsec:simul} for details). The solid lines correspond to the least squares fit (LSF) using a polynomial of the second degree, while the dashed lines correspond to the MFA. Figure~\ref{fig:JDR100} evidenced that accounting for the busbar/nanowire contact resistance is crucial when the junction resistance dominates over the wire resistance. The MFA correctly describes the behavior of the electrical conductivity for $R_\text{j} \gg R_\text{w}$.

\begin{figure}[!htbp]
  \centering
  \includegraphics[width=\columnwidth]{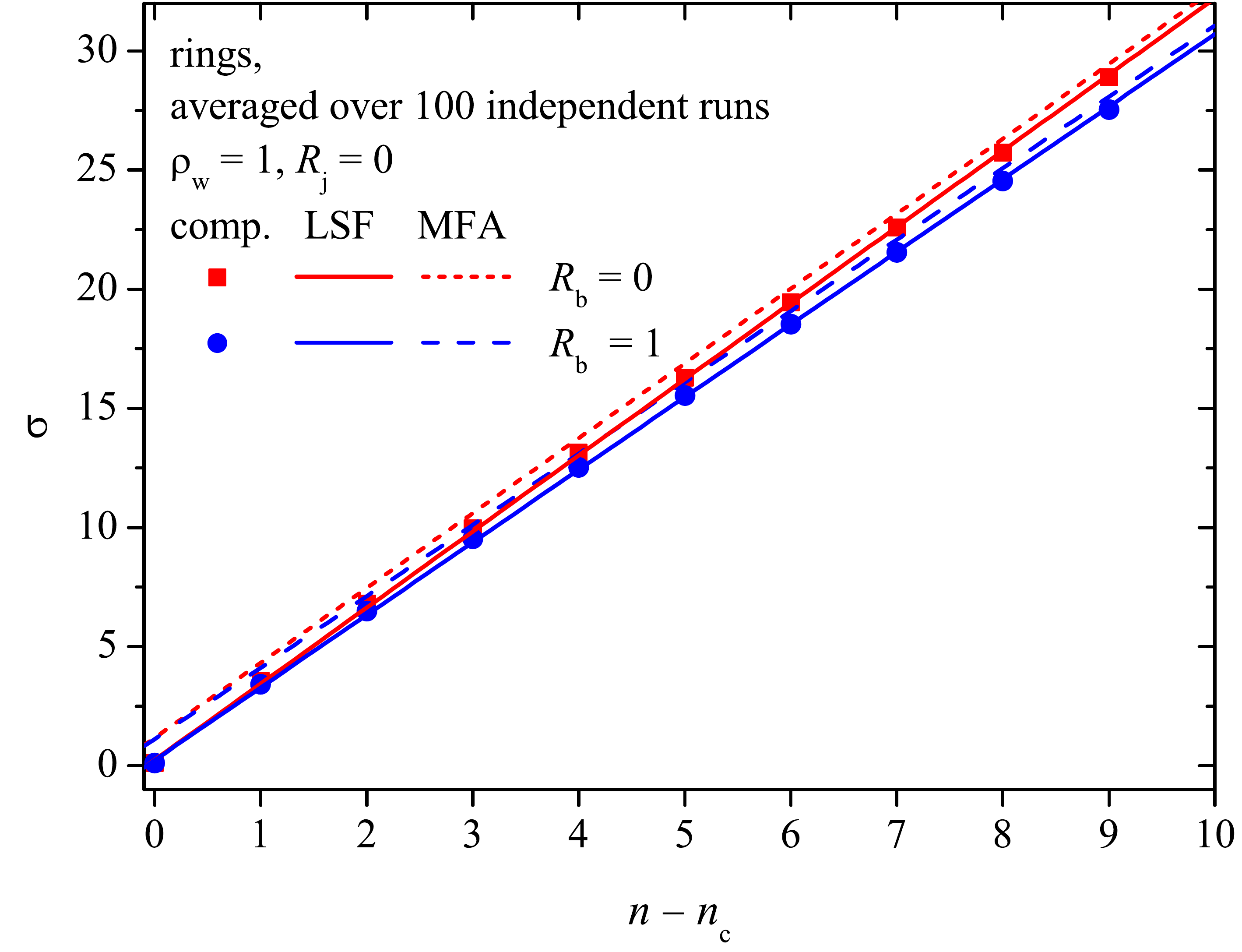}
  \caption{Dependence of the electrical conductance on the number density of the conductive rings, $\rho_\text{w}=1$, $R_\text{j} =0$. The resistances between a conductive ring and a busbar are $R_\text{b} = 0$ and $R_\text{b} = 1$. The markers correspond to the direct computation of the electrical conductance. The solid lines correspond to the least squares approximation using a linear function. The dashed lines correspond to the mean-field approximation (Eq.~\eqref{eq:RefflimWDRbus}).}\label{fig:WDR100}
\end{figure}

Figure~\ref{fig:WDR100} demonstrates the electrical conductance against the number density of the conductive rings when the wire resistance dominates over the junction resistance ($\rho_\text{w}=1$, $R_\text{j} = 0$). In this particular case, Eq.~\eqref{eq:Refflim} leads to the electrical conductance
\begin{equation}\label{eq:RefflimWDRbus}
\sigma_0 = \frac{ 2 \pi n r L }{ 2 \rho_\text{w} L + \pi R_\text{b}}.
\end{equation}
The markers correspond to the direct computation of the electrical conductance, viz., $R_\text{b} = 0$ (squares) and $R_\text{b} = 1$ (circles). The solid lines correspond to the LSF using a linear function. The dashed lines correspond to the MFA. Again, the MFA correctly describes the behavior of the electrical conductance when the wire resistance dominates over the junction resistance. However, the effect of the busbars is much smaller as compared to the previous case.

\begin{figure}[!htbp]
  \centering
  \includegraphics[width=\columnwidth]{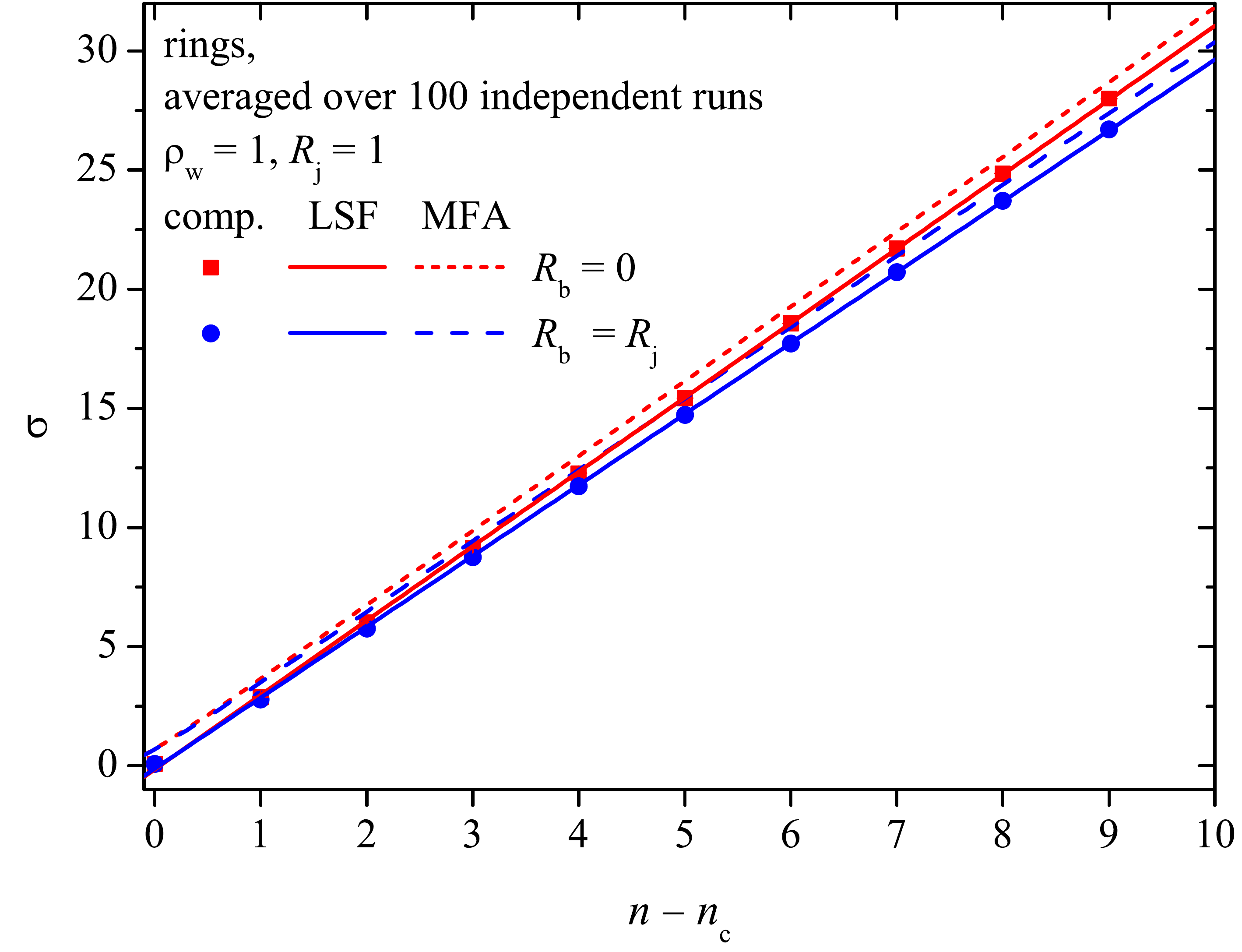}
  \caption{Dependence of the electrical conductance on the number density of the conductive rings, $\rho_\text{w}=1$, $R_\text{j} =1$. The markers correspond to the direct computation of the electrical conductance, viz., $R_\text{b} = 0$ (squares) and $R_\text{b} = 1$ (circles). The solid lines correspond to the least squares approximation using a linear function. The dashed lines correspond to the mean-field approximation. }\label{fig:JWR100}
\end{figure}

Figure~\ref{fig:JWR100} shows the electrical conductance against the number density of the conductive rings when the wire resistance equals the junction resistance ($\rho_\text{w}=1$, $R_\text{j} = 1$). The markers correspond to the direct computation of the electrical conductance, viz., $R_\text{b} = 0$ (squares) and $R_\text{b} = 1$ (circles). The solid lines correspond to the LSF using a linear function. The dashed lines correspond to the MFA. In this common case, $\sigma_0 = R_0^{-1}$, where $ R_0$ is defined by Eq.~\eqref{eq:Refflim}.

To check the contribution of the busbars to the electrical resistance, the dependencies of the resistance on the system size, $L$, as well as on the number density, $n$, have been studied. Formulae~\eqref{eq:Refflim} and~\eqref{eq:RstickJDR} predict a linear dependency of this contribution, both on $n^{-1}$ and on $L^{-1}$.

\begin{figure}[!htbp]
  \centering
  \includegraphics[width=\columnwidth]{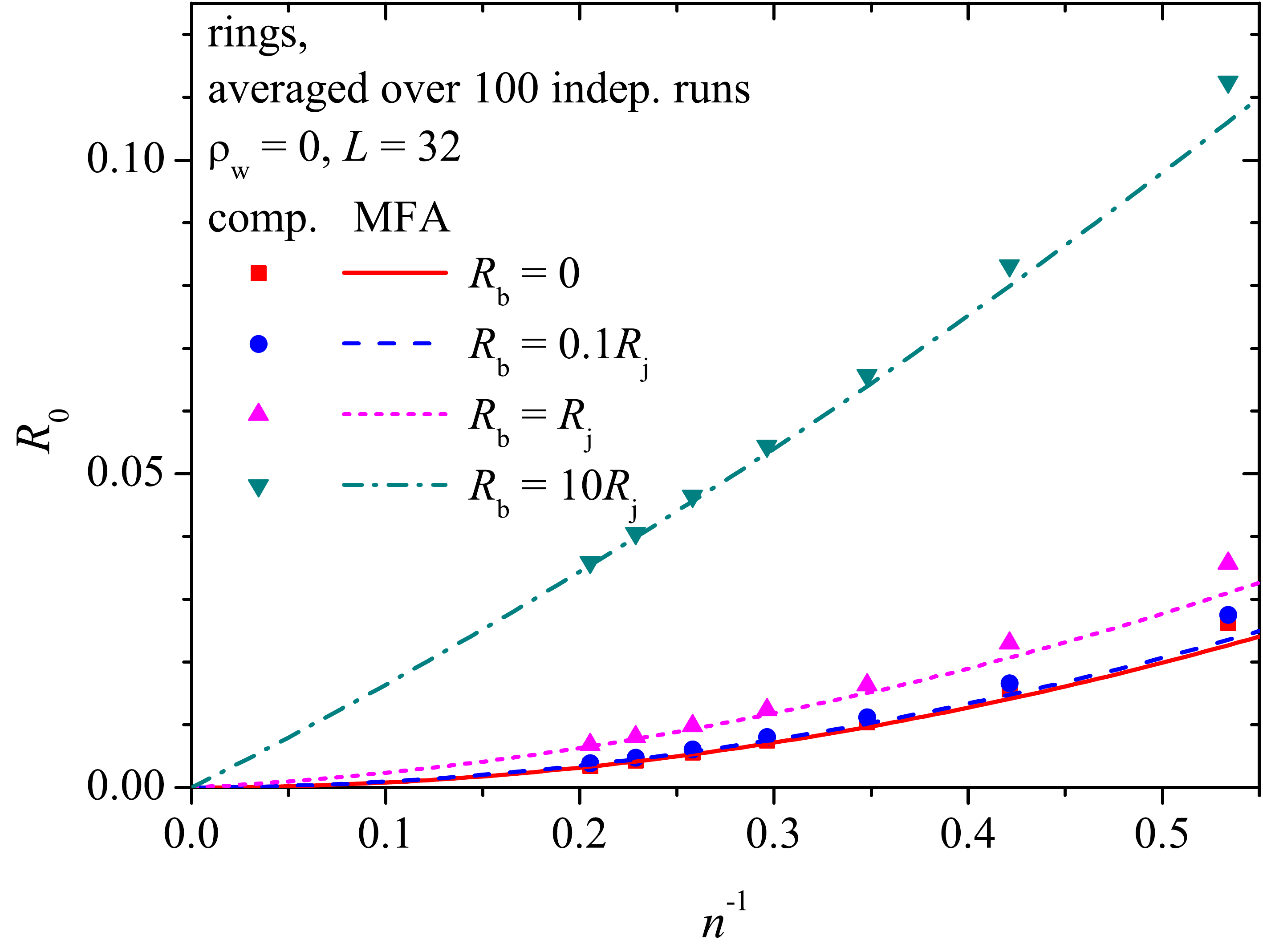}
  \caption{Dependency of the resistance on the reciprocal number density ($n^{-1}$) for different values of the ratio $R_\text{b}/R_\text{j}$ when $\rho_\text{w} = 0$; $L=32$; $R_\text{j} = 1$. The lines correspond to Eq.~\eqref{eq:Refflim}.}\label{fig:R0busesRings}
\end{figure}

Figure~\ref{fig:R0busesRings} demonstrates the electrical resistance against the reciprocal number density $n^{-1}$ for different values of the ratio $R_\text{b}/R_\text{j}$ when $\rho_\text{w} = 0$. The results of the direct computations (markers) are compared with the predictions of the MFA (lines). The larger the number density, the more accurate the prediction, since the contribution of the ring/ring contacts to the electrical resistance decreases rapidly as $n^{-2}$.

\begin{figure}[!htbp]
  \centering
  \includegraphics[width=\columnwidth]{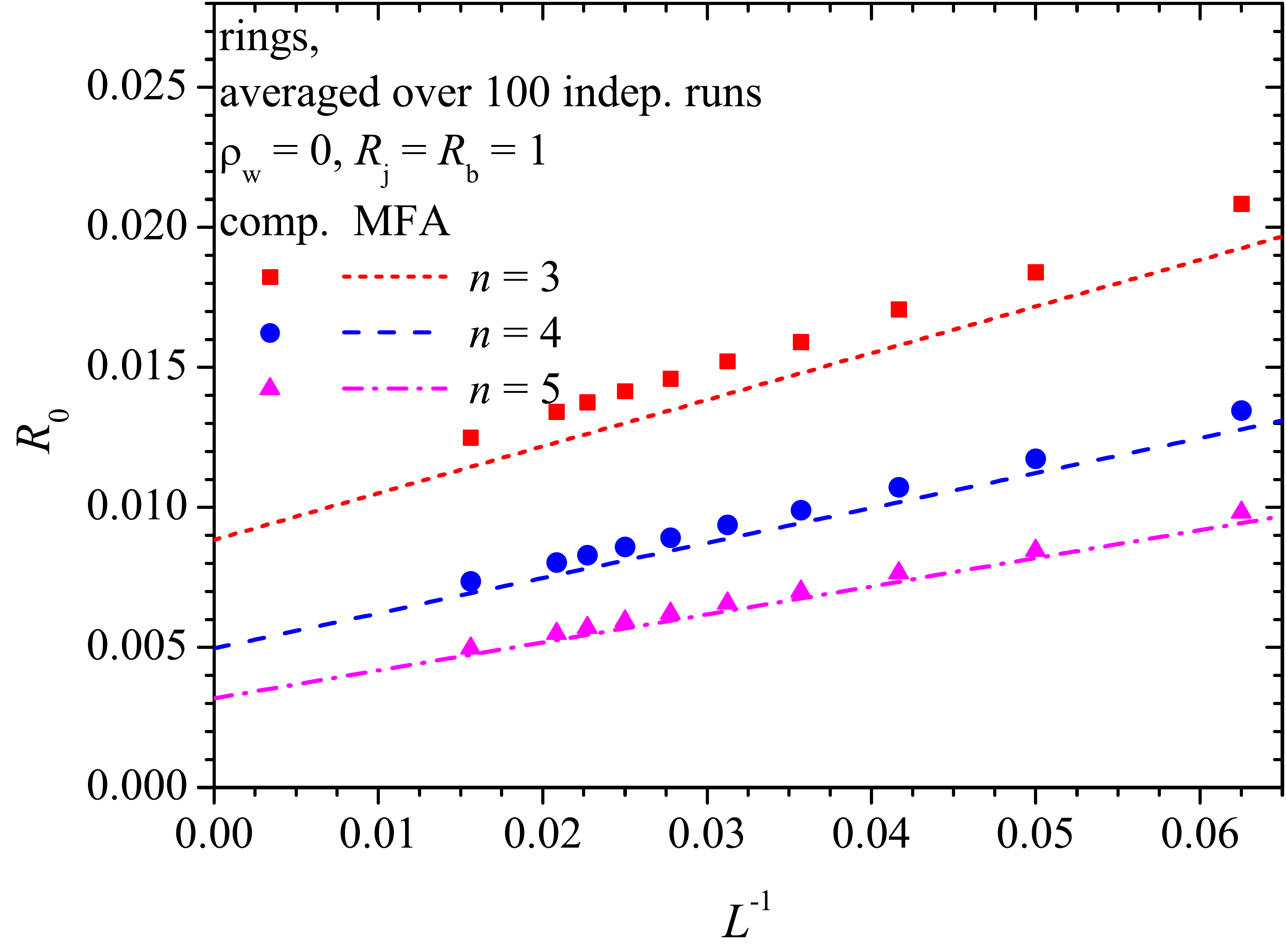}
  \caption{Dependency of the resistance on the reciprocal linear system size $L^{-1}$ for the three different values of the number density when $\rho_\text{w} = 0$ and $R_\text{b} = R_\text{j} = 1$. The lines correspond to Eq.~\eqref{eq:Refflim}.}\label{fig:R0vsLrecip}
\end{figure}

Figure~\ref{fig:R0vsLrecip} presents the resistance against the reciprocal linear size of the system under consideration for the three different values of the number density when $\rho_\text{w} = 0$, $R_\text{b} = R_\text{j} = 1$. The lines correspond to Eq.~\eqref{eq:Refflim}. As the size of the system increases, the contribution of the busbars to the resistance decreases.

\subsection{Stick-based conductive films}\label{subsec:SBCF}
\begin{figure}[!htbp]
  \centering
  \includegraphics[width=\columnwidth]{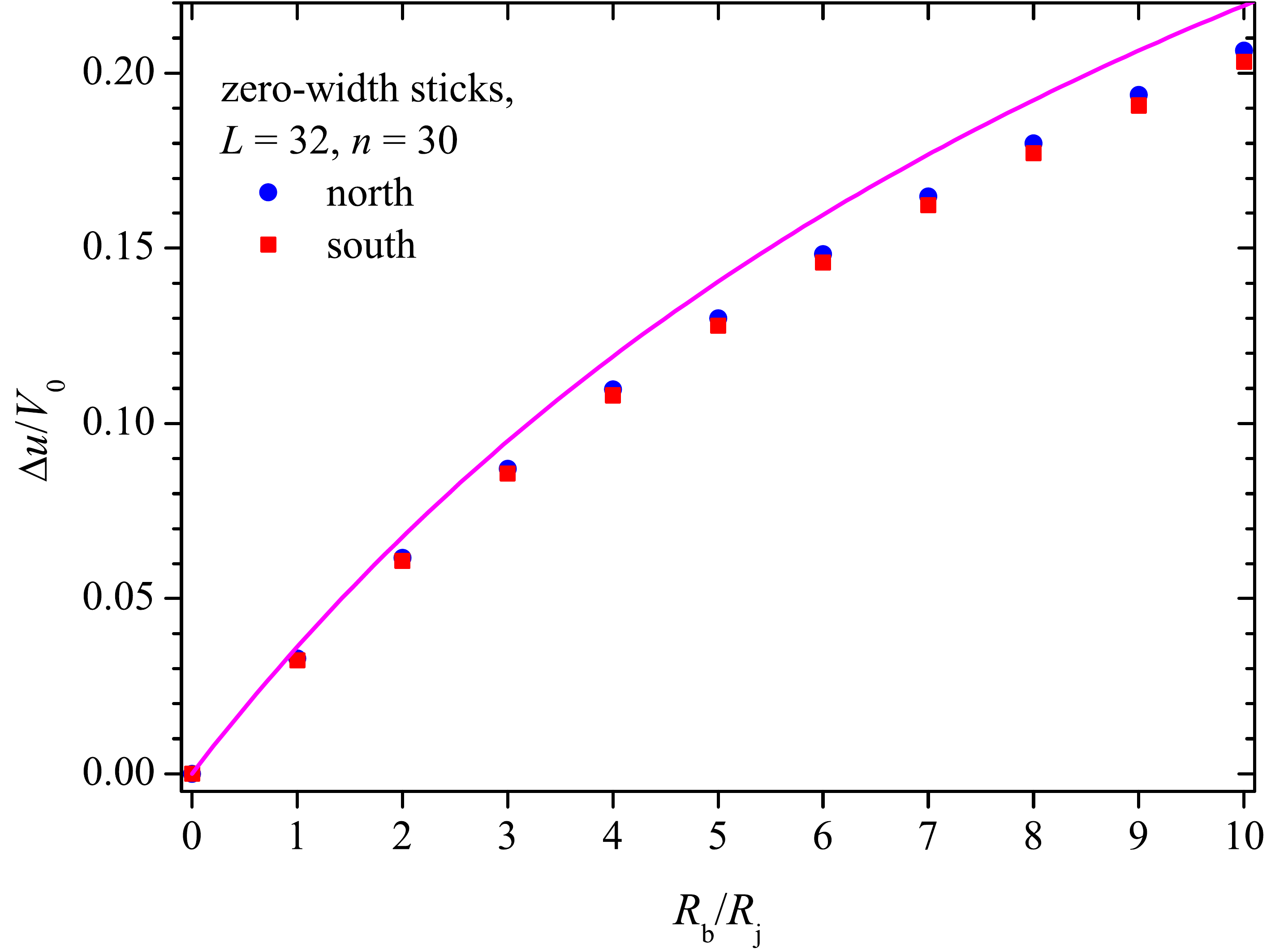}
  \caption{Dependency of the potential jump between the busbar and a conductor on the value of the ratio $R_\text{b} / R_\text{j}$ for systems of randomly placed conductive sticks. $L=32$, $n = 30$, $\rho_\text{w} =0$. ``North'' and ``south'' correspond to the upper and lower busbars. The line corresponds to Eq.~\eqref{eq:potjumpsticks}.}\label{fig:potjumpsticks}
\end{figure}

Figure~\ref{fig:potjumpsticks} shows the dependency of the potential jump between the busbar and a conductor on the value of the ratio $R_\text{b} / R_\text{j}$ for systems of randomly placed conductive sticks. Combination of Eqs.~\eqref{eq:Deltau}, \eqref{eq:RstickJDR}, and~\eqref{eq:Nbstick} with Ohm's law leads to
\begin{equation}\label{eq:potjumpsticks}
  \Delta u = V_0 \frac{z n l^3}{24 L + 2 n l^3 z}, \quad z = R_\text{b} / R_\text{j}.
\end{equation}

\begin{figure}[!htbp]
  \centering
  \includegraphics[width=\columnwidth]{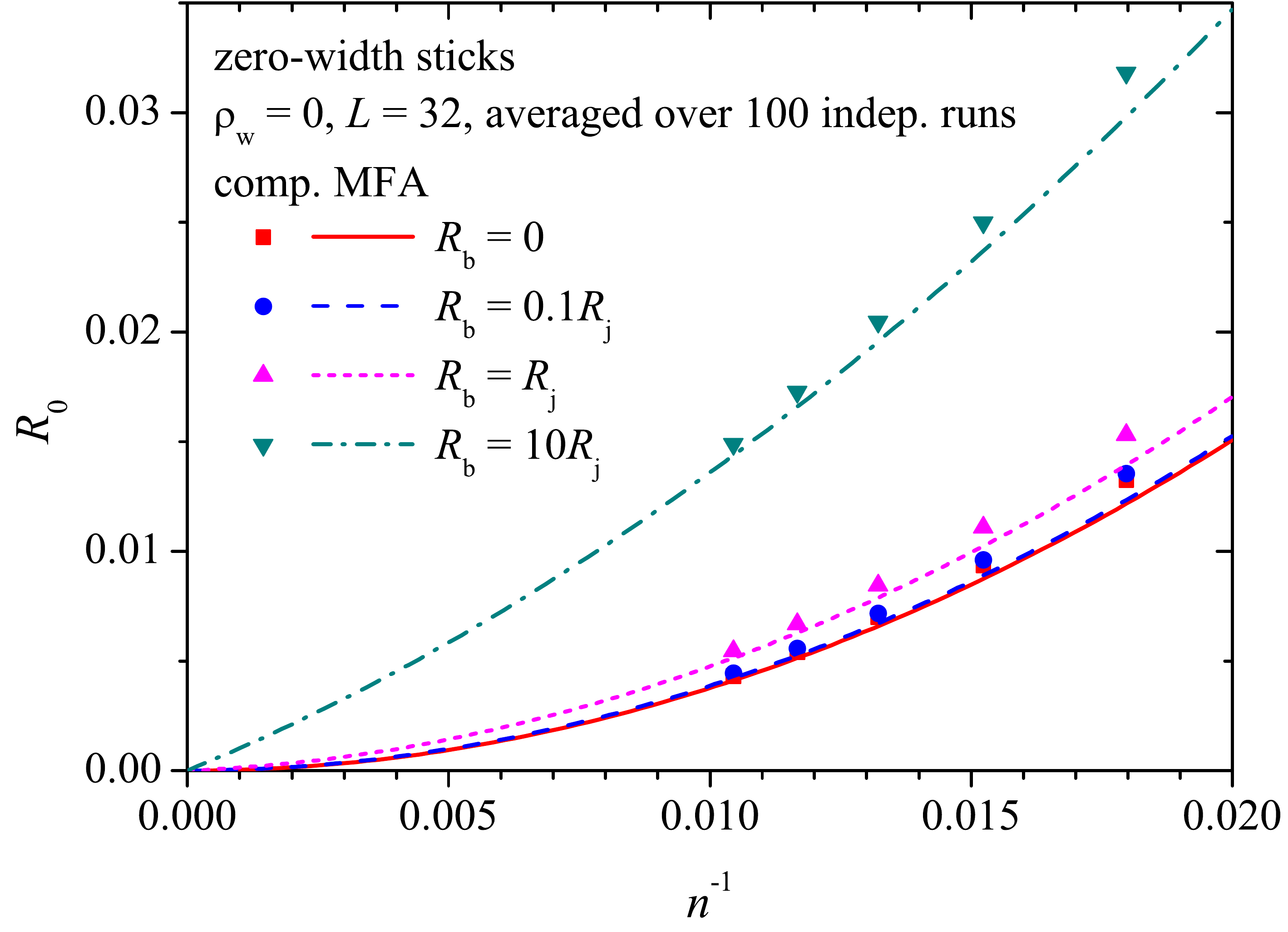}
  \caption{Dependency of the resistance on the reciprocal number density ($n^{-1}$) for different values of the ratio $R_\text{b}/R_\text{j}$ when $\rho_\text{w} = 0$; $L=32$; $R_\text{j} = 1$. The lines correspond to Eq.~\eqref{eq:RstickJDR}.}\label{fig:R0busesSticks}
\end{figure}

Figure~\ref{fig:R0busesSticks} demonstrates the electrical resistance against the reciprocal number density $n^{-1}$ for different values of the ratio $R_\text{b}/R_\text{j}$ when $\rho_\text{w} = 0$. The results of the direct computations (markers) are compared with the predictions of the MFA (lines). The larger the number density, the more accurate the prediction, since the contribution of the stick/stick contacts to the electrical resistance decreases rapidly as $n^{-2}$.

\begin{figure}[!htbp]
  \centering
  \includegraphics[width=\columnwidth]{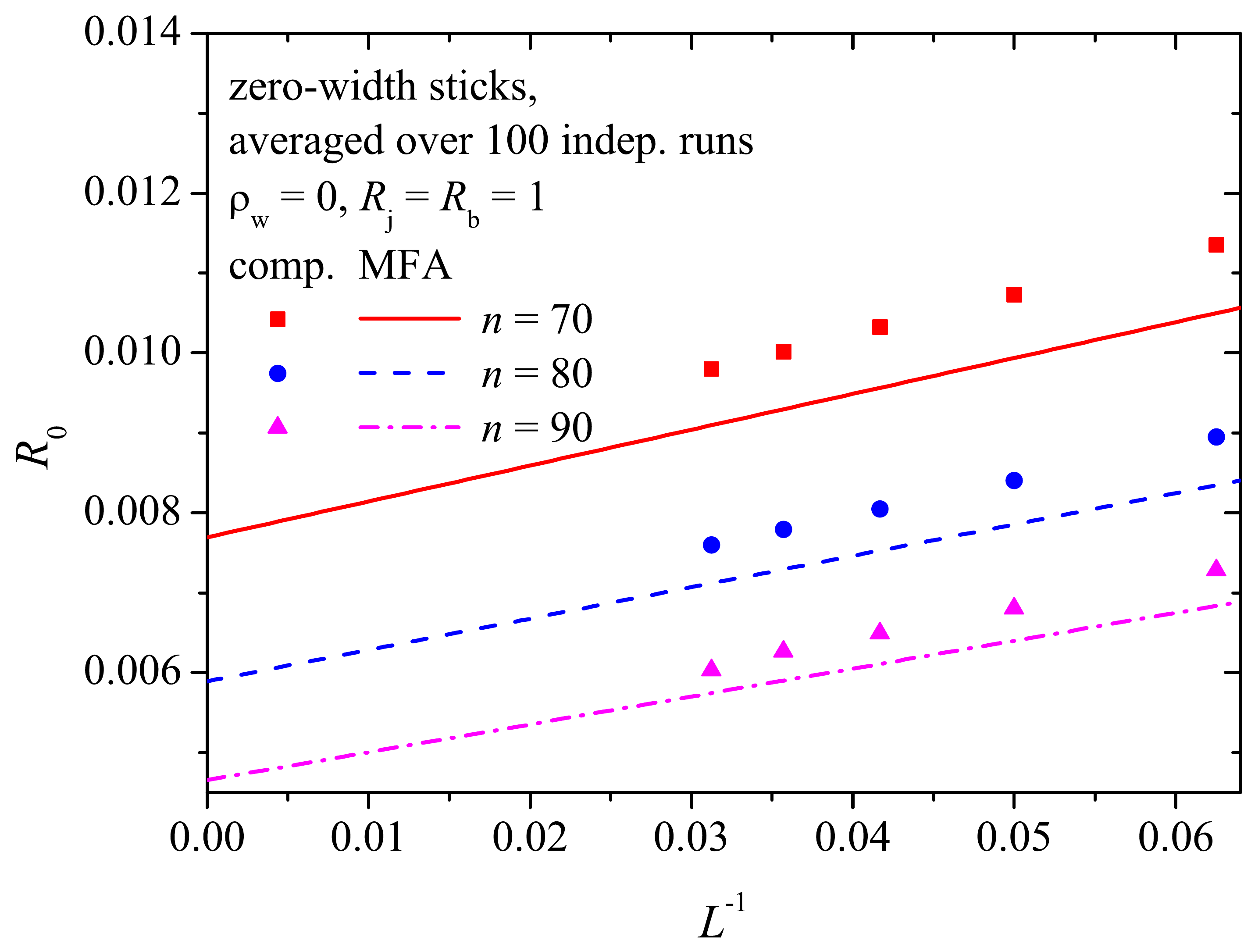}
  \caption{Dependency of the resistance on the reciprocal linear size of the system $L^{-1}$ for the three different values of the number density when $\rho_\text{w} = 0$ and $R_\text{b} = R_\text{j} = 1$. The lines correspond to Eq.~\eqref{eq:RstickJDR}.}\label{fig:R0vsLrecipSticks}
\end{figure}

Figure~\ref{fig:R0vsLrecipSticks} presents the resistance against the reciprocal linear size of the system under consideration for the three different values of the number density when $\rho_\text{w} = 0$, $R_\text{b} = R_\text{j} = 1$. The lines correspond to Eq.~\eqref{eq:RstickJDR}. As the size of the system increases, the contribution of the busbars to the resistance decreases. The results of computer simulations show the same trend as theoretical predictions.

\section{Conclusion}\label{sec:concl}

We found that the busbar/nanowire contact resistance is crucial when the junction resistance, $R_\text{j}$, dominates over the wire resistance, $R_\text{w}$. Accounting for the busbar/nanowire contact resistance leads to a correct description of the behavior of the electrical conductivity for any ratio $R_\text{j}/R_\text{w}$ within the MFA. Recently, a deviation between computations and the MFA predictions in systems of randomly placed conductive rings when $R_\text{j} \gg R_\text{w}$, while $R_\text{j} = R_\text{b}$ has been reported.\cite{Tarasevich2021JAP} In the present study, we have shown that this discrepancy is completely eliminated if, when carrying out a theoretical assessment of the electrical conductivity, the busbar/nanowire contact resistance is taken into account. However, in the thermodynamic limit ($L \to \infty$), the electrical resistance of the system under consideration is independent on the resistance of the contacts of the conductors with the busbars. Comparison of~\eqref{eq:Benda} with~\eqref{eq:Refflim} and~\eqref{eq:split} evidenced that $A \sim N^{-1}(l/L)^{-2}$, $B \sim N^{-2}(l/L)^{-4}$, and $C \sim N^{-1}(l/L)^{-1}$ ($l=r$ in the case of rings). These exponents of $N$ for $A$ and $B$ are consistent with the exponents proposed for large dense networks, well above the percolation threshold.\cite{Zezelj2012PRB} However, these exponents are noticeably different from those obtained for small systems ($L/l=4$).\cite{Benda2019}

In the case of nanorod-based 2D systems, the transition from a continuous consideration in the framework of the MFA\cite{Tarasevich2022PRE} to a discrete one\cite{Tarasevich2022PCCP} somewhat improved the estimates of the electrical conductivity. By contrast, in the case of nanoring-based 2D systems, the discrete consideration in the framework of the MFA, carried out in the present study, gave no improvement as compared to the continuous consideration.\cite{Tarasevich2021JAP} In any case, although the MFA is supposed to be valid only for very dense systems, its estimates of the electrical conductivity are fairly reasonable even at very moderate values of the number density.

\begin{acknowledgments}
The authors (Y.Y.T. and A.V.E.) acknowledge funding from the Foundation for the Advancement of Theoretical Physics and Mathematics ``BASIS'', grant~20-1-1-8-1.
\end{acknowledgments}

\section*{Author declarations}
\subsection*{Conflict of Interest}
The authors have no conflicts to disclose.
%\subsection*{Author Contributions}

\section*{Data availability}
The data that support the findings of this study are available from the corresponding author upon reasonable request.
\bibliography{MFAdiscr}

%aipnum4-2.bst 2019-01-14 (MD) hand-edited version of apsrev4-1.bst
\begin{thebibliography}{33}%
\makeatletter
\providecommand \@ifxundefined [1]{%
 \@ifx{#1\undefined}
}%
\providecommand \@ifnum [1]{%
 \ifnum #1\expandafter \@firstoftwo
 \else \expandafter \@secondoftwo
 \fi
}%
\providecommand \@ifx [1]{%
 \ifx #1\expandafter \@firstoftwo
 \else \expandafter \@secondoftwo
 \fi
}%
\providecommand \natexlab [1]{#1}%
\providecommand \enquote  [1]{``#1''}%
\providecommand \bibnamefont  [1]{#1}%
\providecommand \bibfnamefont [1]{#1}%
\providecommand \citenamefont [1]{#1}%
\providecommand \href@noop [0]{\@secondoftwo}%
\providecommand \href [0]{\begingroup \@sanitize@url \@href}%
\providecommand \@href[1]{\@@startlink{#1}\@@href}%
\providecommand \@@href[1]{\endgroup#1\@@endlink}%
\providecommand \@sanitize@url [0]{\catcode `\\12\catcode `\$12\catcode
  `\&12\catcode `\#12\catcode `\^12\catcode `\_12\catcode `\%12\relax}%
\providecommand \@@startlink[1]{}%
\providecommand \@@endlink[0]{}%
\providecommand \url  [0]{\begingroup\@sanitize@url \@url }%
\providecommand \@url [1]{\endgroup\@href {#1}{\urlprefix }}%
\providecommand \urlprefix  [0]{URL }%
\providecommand \Eprint [0]{\href }%
\providecommand \doibase [0]{https://doi.org/}%
\providecommand \selectlanguage [0]{\@gobble}%
\providecommand \bibinfo  [0]{\@secondoftwo}%
\providecommand \bibfield  [0]{\@secondoftwo}%
\providecommand \translation [1]{[#1]}%
\providecommand \BibitemOpen [0]{}%
\providecommand \bibitemStop [0]{}%
\providecommand \bibitemNoStop [0]{.\EOS\space}%
\providecommand \EOS [0]{\spacefactor3000\relax}%
\providecommand \BibitemShut  [1]{\csname bibitem#1\endcsname}%
\let\auto@bib@innerbib\@empty
%</preamble>
\bibitem [{\citenamefont {Liu}\ \emph {et~al.}(2019)\citenamefont {Liu},
  \citenamefont {Jia}, \citenamefont {Gardner}, \citenamefont {Johansson},\
  and\ \citenamefont {Zhang}}]{Liu2019}%
  \BibitemOpen
  \bibfield  {author} {\bibinfo {author} {\bibfnamefont {J.}~\bibnamefont
  {Liu}}, \bibinfo {author} {\bibfnamefont {D.}~\bibnamefont {Jia}}, \bibinfo
  {author} {\bibfnamefont {J.~M.}\ \bibnamefont {Gardner}}, \bibinfo {author}
  {\bibfnamefont {E.~M.}\ \bibnamefont {Johansson}},\ and\ \bibinfo {author}
  {\bibfnamefont {X.}~\bibnamefont {Zhang}},\ }\bibfield  {title} {\enquote
  {\bibinfo {title} {Metal nanowire networks: Recent advances and challenges
  for new generation photovoltaics},}\ }\href
  {https://doi.org/10.1016/j.mtener.2019.05.007} {\bibfield  {journal}
  {\bibinfo  {journal} {Mater. Today Energy}\ }\textbf {\bibinfo {volume}
  {13}},\ \bibinfo {pages} {152--185} (\bibinfo {year} {2019})}\BibitemShut
  {NoStop}%
\bibitem [{\citenamefont {Benda}, \citenamefont {Canc{\`{e}}s},\ and\
  \citenamefont {Lebental}(2019)}]{Benda2019}%
  \BibitemOpen
  \bibfield  {author} {\bibinfo {author} {\bibfnamefont {R.}~\bibnamefont
  {Benda}}, \bibinfo {author} {\bibfnamefont {E.}~\bibnamefont
  {Canc{\`{e}}s}},\ and\ \bibinfo {author} {\bibfnamefont {B.}~\bibnamefont
  {Lebental}},\ }\bibfield  {title} {\enquote {\bibinfo {title} {Effective
  resistance of random percolating networks of stick nanowires: {Functional}
  dependence on elementary physical parameters},}\ }\href
  {https://doi.org/10.1063/1.5108575} {\bibfield  {journal} {\bibinfo
  {journal} {J. Appl. Phys.}\ }\textbf {\bibinfo {volume} {126}},\ \bibinfo
  {pages} {044306} (\bibinfo {year} {2019})}\BibitemShut {NoStop}%
\bibitem [{\citenamefont {Ponzoni}(2019)}]{Ponzoni2019APL}%
  \BibitemOpen
  \bibfield  {author} {\bibinfo {author} {\bibfnamefont {A.}~\bibnamefont
  {Ponzoni}},\ }\bibfield  {title} {\enquote {\bibinfo {title} {The
  contributions of junctions and nanowires/nanotubes in conductive networks},}\
  }\href {https://doi.org/10.1063/1.5090117} {\bibfield  {journal} {\bibinfo
  {journal} {Appl. Phys. Lett.}\ }\textbf {\bibinfo {volume} {114}},\ \bibinfo
  {pages} {153105} (\bibinfo {year} {2019})}\BibitemShut {NoStop}%
\bibitem [{\citenamefont {Kim}\ and\ \citenamefont {Nam}(2020)}]{Kim2020}%
  \BibitemOpen
  \bibfield  {author} {\bibinfo {author} {\bibfnamefont {D.}~\bibnamefont
  {Kim}}\ and\ \bibinfo {author} {\bibfnamefont {J.}~\bibnamefont {Nam}},\
  }\bibfield  {title} {\enquote {\bibinfo {title} {Electrical conductivity
  analysis for networks of conducting rods using a block matrix approach: {A}
  case study under junction resistance dominant assumption},}\ }\href
  {https://doi.org/10.1021/acs.jpcc.9b07163} {\bibfield  {journal} {\bibinfo
  {journal} {J. Phys. Chem. C}\ }\textbf {\bibinfo {volume} {124}},\ \bibinfo
  {pages} {986--996} (\bibinfo {year} {2020})}\BibitemShut {NoStop}%
\bibitem [{\citenamefont {Fata}\ \emph {et~al.}(2020)\citenamefont {Fata},
  \citenamefont {Mishra}, \citenamefont {Xue}, \citenamefont {Wang},
  \citenamefont {Hicks},\ and\ \citenamefont {Ural}}]{Fata2020JAP}%
  \BibitemOpen
  \bibfield  {author} {\bibinfo {author} {\bibfnamefont {N.}~\bibnamefont
  {Fata}}, \bibinfo {author} {\bibfnamefont {S.}~\bibnamefont {Mishra}},
  \bibinfo {author} {\bibfnamefont {Y.}~\bibnamefont {Xue}}, \bibinfo {author}
  {\bibfnamefont {Y.}~\bibnamefont {Wang}}, \bibinfo {author} {\bibfnamefont
  {J.}~\bibnamefont {Hicks}},\ and\ \bibinfo {author} {\bibfnamefont
  {A.}~\bibnamefont {Ural}},\ }\bibfield  {title} {\enquote {\bibinfo {title}
  {Effect of junction-to-nanowire resistance ratio on the percolation
  conductivity and critical exponents of nanowire networks},}\ }\href
  {https://doi.org/10.1063/5.0023209} {\bibfield  {journal} {\bibinfo
  {journal} {J. Appl. Phys.}\ }\textbf {\bibinfo {volume} {128}},\ \bibinfo
  {pages} {124301} (\bibinfo {year} {2020})}\BibitemShut {NoStop}%
\bibitem [{\citenamefont {Manning}\ \emph {et~al.}(2020)\citenamefont
  {Manning}, \citenamefont {Flowers}, \citenamefont {Cruz}, \citenamefont
  {Gomes~da Rocha}, \citenamefont {O'Callaghan}, \citenamefont {Ferreira},
  \citenamefont {Wiley},\ and\ \citenamefont {Boland}}]{Manning2020}%
  \BibitemOpen
  \bibfield  {author} {\bibinfo {author} {\bibfnamefont {H.~G.}\ \bibnamefont
  {Manning}}, \bibinfo {author} {\bibfnamefont {P.~F.}\ \bibnamefont
  {Flowers}}, \bibinfo {author} {\bibfnamefont {M.~A.}\ \bibnamefont {Cruz}},
  \bibinfo {author} {\bibfnamefont {C.}~\bibnamefont {Gomes~da Rocha}},
  \bibinfo {author} {\bibfnamefont {C.}~\bibnamefont {O'Callaghan}}, \bibinfo
  {author} {\bibfnamefont {M.~S.}\ \bibnamefont {Ferreira}}, \bibinfo {author}
  {\bibfnamefont {B.~J.}\ \bibnamefont {Wiley}},\ and\ \bibinfo {author}
  {\bibfnamefont {J.~J.}\ \bibnamefont {Boland}},\ }\bibfield  {title}
  {\enquote {\bibinfo {title} {The resistance of {Cu} nanowire--nanowire
  junctions and electro-optical modeling of {Cu} nanowire networks},}\ }\href
  {https://doi.org/10.1063/5.0012005} {\bibfield  {journal} {\bibinfo
  {journal} {Appl. Phys. Lett.}\ }\textbf {\bibinfo {volume} {116}},\ \bibinfo
  {pages} {251902} (\bibinfo {year} {2020})}\BibitemShut {NoStop}%
\bibitem [{\citenamefont {Tarasevich}, \citenamefont {Eserkepov},\ and\
  \citenamefont {Vodolazskaya}(2021)}]{Tarasevich2021JAP}%
  \BibitemOpen
  \bibfield  {author} {\bibinfo {author} {\bibfnamefont {Y.~Y.}\ \bibnamefont
  {Tarasevich}}, \bibinfo {author} {\bibfnamefont {A.~V.}\ \bibnamefont
  {Eserkepov}},\ and\ \bibinfo {author} {\bibfnamefont {I.~V.}\ \bibnamefont
  {Vodolazskaya}},\ }\bibfield  {title} {\enquote {\bibinfo {title} {Electrical
  conductivity of nanoring-based transparent conductive films: A mean-field
  approach},}\ }\href {https://doi.org/10.1063/5.0078219} {\bibfield  {journal}
  {\bibinfo  {journal} {J. Appl. Phys.}\ }\textbf {\bibinfo {volume} {130}},\
  \bibinfo {pages} {244302} (\bibinfo {year} {2021})}\BibitemShut {NoStop}%
\bibitem [{\citenamefont {Tarasevich}, \citenamefont {Vodolazskaya},\ and\
  \citenamefont {Eserkepov}(2022)}]{Tarasevich2022PCCP}%
  \BibitemOpen
  \bibfield  {author} {\bibinfo {author} {\bibfnamefont {Y.~Y.}\ \bibnamefont
  {Tarasevich}}, \bibinfo {author} {\bibfnamefont {I.~V.}\ \bibnamefont
  {Vodolazskaya}},\ and\ \bibinfo {author} {\bibfnamefont {A.~V.}\ \bibnamefont
  {Eserkepov}},\ }\bibfield  {title} {\enquote {\bibinfo {title} {Electrical
  conductivity of random metallic nanowire networks: an analytical
  consideration along with computer simulation},}\ }\href
  {https://doi.org/10.1039/D2CP00936F} {\bibfield  {journal} {\bibinfo
  {journal} {Phys. Chem. Chem. Phys.}\ }\textbf {\bibinfo {volume} {24}},\
  \bibinfo {pages} {11812--11819} (\bibinfo {year} {2022})}\BibitemShut
  {NoStop}%
\bibitem [{\citenamefont {Tarasevich}, \citenamefont {Eserkepov},\ and\
  \citenamefont {Vodolazskaya}(2022)}]{Tarasevich2022PRE}%
  \BibitemOpen
  \bibfield  {author} {\bibinfo {author} {\bibfnamefont {Y.~Y.}\ \bibnamefont
  {Tarasevich}}, \bibinfo {author} {\bibfnamefont {A.~V.}\ \bibnamefont
  {Eserkepov}},\ and\ \bibinfo {author} {\bibfnamefont {I.~V.}\ \bibnamefont
  {Vodolazskaya}},\ }\bibfield  {title} {\enquote {\bibinfo {title} {Electrical
  conductivity of nanorod-based transparent electrodes: Comparison of
  mean-field approaches},}\ }\href
  {https://doi.org/10.1103/PhysRevE.105.044129} {\bibfield  {journal} {\bibinfo
   {journal} {Phys. Rev. E}\ }\textbf {\bibinfo {volume} {105}},\ \bibinfo
  {pages} {044129} (\bibinfo {year} {2022})}\BibitemShut {NoStop}%
\bibitem [{\citenamefont {{\v{Z}}e{\v{z}}elj}\ and\ \citenamefont
  {Stankovi{\'{c}}}(2012)}]{Zezelj2012PRB}%
  \BibitemOpen
  \bibfield  {author} {\bibinfo {author} {\bibfnamefont {M.}~\bibnamefont
  {{\v{Z}}e{\v{z}}elj}}\ and\ \bibinfo {author} {\bibfnamefont
  {I.}~\bibnamefont {Stankovi{\'{c}}}},\ }\bibfield  {title} {\enquote
  {\bibinfo {title} {From percolating to dense random stick networks:
  {Conductivity} model investigation},}\ }\href
  {https://doi.org/10.1103/physrevb.86.134202} {\bibfield  {journal} {\bibinfo
  {journal} {Phys. Rev. B}\ }\textbf {\bibinfo {volume} {86}},\ \bibinfo
  {pages} {134202} (\bibinfo {year} {2012})}\BibitemShut {NoStop}%
\bibitem [{\citenamefont {Kumar}, \citenamefont {Vidhyadhiraja},\ and\
  \citenamefont {Kulkarni}(2017)}]{Kumar2017JAP}%
  \BibitemOpen
  \bibfield  {author} {\bibinfo {author} {\bibfnamefont {A.}~\bibnamefont
  {Kumar}}, \bibinfo {author} {\bibfnamefont {N.~S.}\ \bibnamefont
  {Vidhyadhiraja}},\ and\ \bibinfo {author} {\bibfnamefont {G.~U.}\
  \bibnamefont {Kulkarni}},\ }\bibfield  {title} {\enquote {\bibinfo {title}
  {Current distribution in conducting nanowire networks},}\ }\href
  {https://doi.org/10.1063/1.4985792} {\bibfield  {journal} {\bibinfo
  {journal} {J. Appl. Phys.}\ }\textbf {\bibinfo {volume} {122}},\ \bibinfo
  {pages} {045101} (\bibinfo {year} {2017})}\BibitemShut {NoStop}%
\bibitem [{\citenamefont {Ainsworth}, \citenamefont {Derby},\ and\
  \citenamefont {Sampson}(2018)}]{Ainsworth2018}%
  \BibitemOpen
  \bibfield  {author} {\bibinfo {author} {\bibfnamefont {C.~A.}\ \bibnamefont
  {Ainsworth}}, \bibinfo {author} {\bibfnamefont {B.}~\bibnamefont {Derby}},\
  and\ \bibinfo {author} {\bibfnamefont {W.~W.}\ \bibnamefont {Sampson}},\
  }\bibfield  {title} {\enquote {\bibinfo {title} {Interdependence of
  resistance and optical transmission in conductive nanowire networks},}\
  }\href {https://doi.org/10.1002/adts.201700011} {\bibfield  {journal}
  {\bibinfo  {journal} {Adv. Theory Simul.}\ }\textbf {\bibinfo {volume} {1}},\
  \bibinfo {pages} {1700011} (\bibinfo {year} {2018})}\BibitemShut {NoStop}%
\bibitem [{\citenamefont {Forr\'{o}}\ \emph {et~al.}(2018)\citenamefont
  {Forr\'{o}}, \citenamefont {Demk\'{o}}, \citenamefont {Weydert},
  \citenamefont {V\"{o}r\"{o}s},\ and\ \citenamefont
  {Tybrandt}}]{Forro2018ACSN}%
  \BibitemOpen
  \bibfield  {author} {\bibinfo {author} {\bibfnamefont {C.}~\bibnamefont
  {Forr\'{o}}}, \bibinfo {author} {\bibfnamefont {L.}~\bibnamefont
  {Demk\'{o}}}, \bibinfo {author} {\bibfnamefont {S.}~\bibnamefont {Weydert}},
  \bibinfo {author} {\bibfnamefont {J.}~\bibnamefont {V\"{o}r\"{o}s}},\ and\
  \bibinfo {author} {\bibfnamefont {K.}~\bibnamefont {Tybrandt}},\ }\bibfield
  {title} {\enquote {\bibinfo {title} {Predictive model for the electrical
  transport within nanowire networks},}\ }\href
  {https://doi.org/10.1021/acsnano.8b05406} {\bibfield  {journal} {\bibinfo
  {journal} {ACS Nano}\ }\textbf {\bibinfo {volume} {12}},\ \bibinfo {pages}
  {11080--11087} (\bibinfo {year} {2018})}\BibitemShut {NoStop}%
\bibitem [{\citenamefont {Fairfield}\ \emph {et~al.}(2014)\citenamefont
  {Fairfield}, \citenamefont {Ritter}, \citenamefont {Bellew}, \citenamefont
  {McCarthy}, \citenamefont {Ferreira},\ and\ \citenamefont
  {Boland}}]{Fairfield2014}%
  \BibitemOpen
  \bibfield  {author} {\bibinfo {author} {\bibfnamefont {J.~A.}\ \bibnamefont
  {Fairfield}}, \bibinfo {author} {\bibfnamefont {C.}~\bibnamefont {Ritter}},
  \bibinfo {author} {\bibfnamefont {A.~T.}\ \bibnamefont {Bellew}}, \bibinfo
  {author} {\bibfnamefont {E.~K.}\ \bibnamefont {McCarthy}}, \bibinfo {author}
  {\bibfnamefont {M.~S.}\ \bibnamefont {Ferreira}},\ and\ \bibinfo {author}
  {\bibfnamefont {J.~J.}\ \bibnamefont {Boland}},\ }\bibfield  {title}
  {\enquote {\bibinfo {title} {Effective electrode length enhances electrical
  activation of nanowire networks: Experiment and simulation},}\ }\href
  {https://doi.org/10.1021/nn5038515} {\bibfield  {journal} {\bibinfo
  {journal} {{ACS} Nano}\ }\textbf {\bibinfo {volume} {8}},\ \bibinfo {pages}
  {9542--9549} (\bibinfo {year} {2014})}\BibitemShut {NoStop}%
\bibitem [{\citenamefont {Kou}\ \emph {et~al.}(2017)\citenamefont {Kou},
  \citenamefont {Yang}, \citenamefont {Chang},\ and\ \citenamefont
  {He}}]{Kou2017}%
  \BibitemOpen
  \bibfield  {author} {\bibinfo {author} {\bibfnamefont {P.}~\bibnamefont
  {Kou}}, \bibinfo {author} {\bibfnamefont {L.}~\bibnamefont {Yang}}, \bibinfo
  {author} {\bibfnamefont {C.}~\bibnamefont {Chang}},\ and\ \bibinfo {author}
  {\bibfnamefont {S.}~\bibnamefont {He}},\ }\bibfield  {title} {\enquote
  {\bibinfo {title} {Improved flexible transparent conductive electrodes based
  on silver nanowire networks by a simple sunlight illumination approach},}\
  }\href {https://doi.org/10.1038/srep42052} {\bibfield  {journal} {\bibinfo
  {journal} {Sci. Rep.}\ }\textbf {\bibinfo {volume} {7}},\ \bibinfo {pages}
  {42052} (\bibinfo {year} {2017})}\BibitemShut {NoStop}%
\bibitem [{\citenamefont {Park}\ \emph {et~al.}(2016)\citenamefont {Park},
  \citenamefont {Hwang}, \citenamefont {Kim}, \citenamefont {Seo},
  \citenamefont {Park}, \citenamefont {Yu}, \citenamefont {Kim},\ and\
  \citenamefont {Lee}}]{Park2016}%
  \BibitemOpen
  \bibfield  {author} {\bibinfo {author} {\bibfnamefont {J.~H.}\ \bibnamefont
  {Park}}, \bibinfo {author} {\bibfnamefont {G.-T.}\ \bibnamefont {Hwang}},
  \bibinfo {author} {\bibfnamefont {S.}~\bibnamefont {Kim}}, \bibinfo {author}
  {\bibfnamefont {J.}~\bibnamefont {Seo}}, \bibinfo {author} {\bibfnamefont
  {H.-J.}\ \bibnamefont {Park}}, \bibinfo {author} {\bibfnamefont
  {K.}~\bibnamefont {Yu}}, \bibinfo {author} {\bibfnamefont {T.-S.}\
  \bibnamefont {Kim}},\ and\ \bibinfo {author} {\bibfnamefont {K.~J.}\
  \bibnamefont {Lee}},\ }\bibfield  {title} {\enquote {\bibinfo {title}
  {Flash-induced self-limited plasmonic welding of silver nanowire network for
  transparent flexible energy harvester},}\ }\href
  {https://doi.org/10.1002/adma.201603473} {\bibfield  {journal} {\bibinfo
  {journal} {Adv. Mater.}\ }\textbf {\bibinfo {volume} {29}},\ \bibinfo {pages}
  {1603473} (\bibinfo {year} {2016})}\BibitemShut {NoStop}%
\bibitem [{\citenamefont {Gomes~da Rocha}\ \emph {et~al.}(2015)\citenamefont
  {Gomes~da Rocha}, \citenamefont {Manning}, \citenamefont {O'Callaghan},
  \citenamefont {Ritter}, \citenamefont {Bellew}, \citenamefont {Boland},\ and\
  \citenamefont {Ferreira}}]{Rocha2015}%
  \BibitemOpen
  \bibfield  {author} {\bibinfo {author} {\bibfnamefont {C.}~\bibnamefont
  {Gomes~da Rocha}}, \bibinfo {author} {\bibfnamefont {H.~G.}\ \bibnamefont
  {Manning}}, \bibinfo {author} {\bibfnamefont {C.}~\bibnamefont
  {O'Callaghan}}, \bibinfo {author} {\bibfnamefont {C.}~\bibnamefont {Ritter}},
  \bibinfo {author} {\bibfnamefont {A.~T.}\ \bibnamefont {Bellew}}, \bibinfo
  {author} {\bibfnamefont {J.~J.}\ \bibnamefont {Boland}},\ and\ \bibinfo
  {author} {\bibfnamefont {M.~S.}\ \bibnamefont {Ferreira}},\ }\bibfield
  {title} {\enquote {\bibinfo {title} {Ultimate conductivity performance in
  metallic nanowire networks},}\ }\href {https://doi.org/10.1039/c5nr03905c}
  {\bibfield  {journal} {\bibinfo  {journal} {Nanoscale}\ }\textbf {\bibinfo
  {volume} {7}},\ \bibinfo {pages} {13011--13016} (\bibinfo {year}
  {2015})}\BibitemShut {NoStop}%
\bibitem [{\citenamefont {Bellew}\ \emph {et~al.}(2015)\citenamefont {Bellew},
  \citenamefont {Manning}, \citenamefont {Gomes~da Rocha}, \citenamefont
  {Ferreira},\ and\ \citenamefont {Boland}}]{Bellew2015}%
  \BibitemOpen
  \bibfield  {author} {\bibinfo {author} {\bibfnamefont {A.~T.}\ \bibnamefont
  {Bellew}}, \bibinfo {author} {\bibfnamefont {H.~G.}\ \bibnamefont {Manning}},
  \bibinfo {author} {\bibfnamefont {C.}~\bibnamefont {Gomes~da Rocha}},
  \bibinfo {author} {\bibfnamefont {M.~S.}\ \bibnamefont {Ferreira}},\ and\
  \bibinfo {author} {\bibfnamefont {J.~J.}\ \bibnamefont {Boland}},\ }\bibfield
   {title} {\enquote {\bibinfo {title} {Resistance of single {Ag} nanowire
  junctions and their role in the conductivity of nanowire networks},}\ }\href
  {https://doi.org/10.1021/acsnano.5b05469} {\bibfield  {journal} {\bibinfo
  {journal} {{ACS} Nano}\ }\textbf {\bibinfo {volume} {9}},\ \bibinfo {pages}
  {11422--11429} (\bibinfo {year} {2015})},\ \bibinfo {note} {supporting info
  DOI: 10.1021/acsnano.5b05469}\BibitemShut {NoStop}%
\bibitem [{\citenamefont {Selzer}\ \emph {et~al.}(2016)\citenamefont {Selzer},
  \citenamefont {Floresca}, \citenamefont {Kneppe}, \citenamefont {Bormann},
  \citenamefont {Sachse}, \citenamefont {Wei{\ss}}, \citenamefont
  {Eychm{\"u}ller}, \citenamefont {Amassian}, \citenamefont
  {M{\"u}ller-Meskamp},\ and\ \citenamefont {Leo}}]{Selzer2016}%
  \BibitemOpen
  \bibfield  {author} {\bibinfo {author} {\bibfnamefont {F.}~\bibnamefont
  {Selzer}}, \bibinfo {author} {\bibfnamefont {C.}~\bibnamefont {Floresca}},
  \bibinfo {author} {\bibfnamefont {D.}~\bibnamefont {Kneppe}}, \bibinfo
  {author} {\bibfnamefont {L.}~\bibnamefont {Bormann}}, \bibinfo {author}
  {\bibfnamefont {C.}~\bibnamefont {Sachse}}, \bibinfo {author} {\bibfnamefont
  {N.}~\bibnamefont {Wei{\ss}}}, \bibinfo {author} {\bibfnamefont
  {A.}~\bibnamefont {Eychm{\"u}ller}}, \bibinfo {author} {\bibfnamefont
  {A.}~\bibnamefont {Amassian}}, \bibinfo {author} {\bibfnamefont
  {L.}~\bibnamefont {M{\"u}ller-Meskamp}},\ and\ \bibinfo {author}
  {\bibfnamefont {K.}~\bibnamefont {Leo}},\ }\bibfield  {title} {\enquote
  {\bibinfo {title} {Electrical limit of silver nanowire electrodes: {Direct}
  measurement of the nanowire junction resistance},}\ }\href
  {https://doi.org/10.1063/1.4947285} {\bibfield  {journal} {\bibinfo
  {journal} {Appl. Phys. Lett.}\ }\textbf {\bibinfo {volume} {108}},\ \bibinfo
  {pages} {163302} (\bibinfo {year} {2016})}\BibitemShut {NoStop}%
\bibitem [{\citenamefont {Lee}\ \emph {et~al.}(2008)\citenamefont {Lee},
  \citenamefont {Connor}, \citenamefont {Cui},\ and\ \citenamefont
  {Peumans}}]{Lee2008}%
  \BibitemOpen
  \bibfield  {author} {\bibinfo {author} {\bibfnamefont {J.-Y.}\ \bibnamefont
  {Lee}}, \bibinfo {author} {\bibfnamefont {S.~T.}\ \bibnamefont {Connor}},
  \bibinfo {author} {\bibfnamefont {Y.}~\bibnamefont {Cui}},\ and\ \bibinfo
  {author} {\bibfnamefont {P.}~\bibnamefont {Peumans}},\ }\bibfield  {title}
  {\enquote {\bibinfo {title} {Solution-processed metal nanowire mesh
  transparent electrodes},}\ }\href {https://doi.org/10.1021/nl073296g}
  {\bibfield  {journal} {\bibinfo  {journal} {Nano Lett.}\ }\textbf {\bibinfo
  {volume} {8}},\ \bibinfo {pages} {689--692} (\bibinfo {year}
  {2008})}\BibitemShut {NoStop}%
\bibitem [{\citenamefont {Nguyen}\ \emph {et~al.}(2019)\citenamefont {Nguyen},
  \citenamefont {Resende}, \citenamefont {Papanastasiou}, \citenamefont
  {Fontanals}, \citenamefont {Jim{\'e}nez}, \citenamefont {Mu{\~n}oz-Rojas},\
  and\ \citenamefont {Bellet}}]{Nguyen2019}%
  \BibitemOpen
  \bibfield  {author} {\bibinfo {author} {\bibfnamefont {V.~H.}\ \bibnamefont
  {Nguyen}}, \bibinfo {author} {\bibfnamefont {J.}~\bibnamefont {Resende}},
  \bibinfo {author} {\bibfnamefont {D.~T.}\ \bibnamefont {Papanastasiou}},
  \bibinfo {author} {\bibfnamefont {N.}~\bibnamefont {Fontanals}}, \bibinfo
  {author} {\bibfnamefont {C.}~\bibnamefont {Jim{\'e}nez}}, \bibinfo {author}
  {\bibfnamefont {D.}~\bibnamefont {Mu{\~n}oz-Rojas}},\ and\ \bibinfo {author}
  {\bibfnamefont {D.}~\bibnamefont {Bellet}},\ }\bibfield  {title} {\enquote
  {\bibinfo {title} {Low-cost fabrication of flexible transparent electrodes
  based on {Al} doped {ZnO} and silver nanowire nanocomposites: impact of the
  network density},}\ }\href {https://doi.org/10.1039/C9NR02664A} {\bibfield
  {journal} {\bibinfo  {journal} {Nanoscale}\ }\textbf {\bibinfo {volume}
  {11}},\ \bibinfo {pages} {12097--12107} (\bibinfo {year} {2019})}\BibitemShut
  {NoStop}%
\bibitem [{\citenamefont {Khanarian}\ \emph {et~al.}(2013)\citenamefont
  {Khanarian}, \citenamefont {Joo}, \citenamefont {Liu}, \citenamefont
  {Eastman}, \citenamefont {Werner}, \citenamefont {O'Connell},\ and\
  \citenamefont {Trefonas}}]{Khanarian2013}%
  \BibitemOpen
  \bibfield  {author} {\bibinfo {author} {\bibfnamefont {G.}~\bibnamefont
  {Khanarian}}, \bibinfo {author} {\bibfnamefont {J.}~\bibnamefont {Joo}},
  \bibinfo {author} {\bibfnamefont {X.-Q.}\ \bibnamefont {Liu}}, \bibinfo
  {author} {\bibfnamefont {P.}~\bibnamefont {Eastman}}, \bibinfo {author}
  {\bibfnamefont {D.}~\bibnamefont {Werner}}, \bibinfo {author} {\bibfnamefont
  {K.}~\bibnamefont {O'Connell}},\ and\ \bibinfo {author} {\bibfnamefont
  {P.}~\bibnamefont {Trefonas}},\ }\bibfield  {title} {\enquote {\bibinfo
  {title} {The optical and electrical properties of silver nanowire mesh
  films},}\ }\href {https://doi.org/10.1063/1.4812390} {\bibfield  {journal}
  {\bibinfo  {journal} {J. Appl. Phys.}\ }\textbf {\bibinfo {volume} {114}},\
  \bibinfo {pages} {024302} (\bibinfo {year} {2013})}\BibitemShut {NoStop}%
\bibitem [{\citenamefont {He}\ \emph {et~al.}(2018)\citenamefont {He},
  \citenamefont {Xu}, \citenamefont {Qiu}, \citenamefont {He},\ and\
  \citenamefont {Zhou}}]{He2018}%
  \BibitemOpen
  \bibfield  {author} {\bibinfo {author} {\bibfnamefont {S.}~\bibnamefont
  {He}}, \bibinfo {author} {\bibfnamefont {X.}~\bibnamefont {Xu}}, \bibinfo
  {author} {\bibfnamefont {X.}~\bibnamefont {Qiu}}, \bibinfo {author}
  {\bibfnamefont {Y.}~\bibnamefont {He}},\ and\ \bibinfo {author}
  {\bibfnamefont {C.}~\bibnamefont {Zhou}},\ }\bibfield  {title} {\enquote
  {\bibinfo {title} {Conductivity of two-dimensional disordered nanowire
  networks: Dependence on length-ratio of conducting paths to all nanowires},}\
  }\href {https://doi.org/10.1063/1.5045176} {\bibfield  {journal} {\bibinfo
  {journal} {J. Appl. Phys.}\ }\textbf {\bibinfo {volume} {124}},\ \bibinfo
  {pages} {054302} (\bibinfo {year} {2018})}\BibitemShut {NoStop}%
\bibitem [{\citenamefont {Xu}\ \emph {et~al.}(2018)\citenamefont {Xu},
  \citenamefont {Xu}, \citenamefont {Mao}, \citenamefont {Shen}, \citenamefont
  {Yu}, \citenamefont {Tan},\ and\ \citenamefont {Song}}]{Xu2018}%
  \BibitemOpen
  \bibfield  {author} {\bibinfo {author} {\bibfnamefont {F.}~\bibnamefont
  {Xu}}, \bibinfo {author} {\bibfnamefont {W.}~\bibnamefont {Xu}}, \bibinfo
  {author} {\bibfnamefont {B.}~\bibnamefont {Mao}}, \bibinfo {author}
  {\bibfnamefont {W.}~\bibnamefont {Shen}}, \bibinfo {author} {\bibfnamefont
  {Y.}~\bibnamefont {Yu}}, \bibinfo {author} {\bibfnamefont {R.}~\bibnamefont
  {Tan}},\ and\ \bibinfo {author} {\bibfnamefont {W.}~\bibnamefont {Song}},\
  }\bibfield  {title} {\enquote {\bibinfo {title} {Preparation and cold welding
  of silver nanowire based transparent electrodes with optical transmittances
  $>90$\% and sheet resistances $<10 $ ohm/sq},}\ }\href
  {https://doi.org/10.1016/j.jcis.2017.10.051} {\bibfield  {journal} {\bibinfo
  {journal} {J. Coll. Interf. Sci.}\ }\textbf {\bibinfo {volume} {512}},\
  \bibinfo {pages} {208--218} (\bibinfo {year} {2018})}\BibitemShut {NoStop}%
\bibitem [{\citenamefont {Oh}\ \emph {et~al.}(2018)\citenamefont {Oh},
  \citenamefont {Oh}, \citenamefont {Kim},\ and\ \citenamefont
  {Yeom}}]{Oh2018}%
  \BibitemOpen
  \bibfield  {author} {\bibinfo {author} {\bibfnamefont {J.~S.}\ \bibnamefont
  {Oh}}, \bibinfo {author} {\bibfnamefont {J.~S.}\ \bibnamefont {Oh}}, \bibinfo
  {author} {\bibfnamefont {T.~H.}\ \bibnamefont {Kim}},\ and\ \bibinfo {author}
  {\bibfnamefont {G.~Y.}\ \bibnamefont {Yeom}},\ }\bibfield  {title} {\enquote
  {\bibinfo {title} {Efficient metallic nanowire welding using the eddy current
  method},}\ }\href {https://doi.org/10.1088/1361-6528/aaf13d} {\bibfield
  {journal} {\bibinfo  {journal} {Nanotechnology}\ }\textbf {\bibinfo {volume}
  {30}},\ \bibinfo {pages} {065708} (\bibinfo {year} {2018})}\BibitemShut
  {NoStop}%
\bibitem [{\citenamefont {Lee}\ \emph {et~al.}(2016)\citenamefont {Lee},
  \citenamefont {Kim}, \citenamefont {Hwang}, \citenamefont {Choi},\ and\
  \citenamefont {Kim}}]{Lee2016}%
  \BibitemOpen
  \bibfield  {author} {\bibinfo {author} {\bibfnamefont {E.-J.}\ \bibnamefont
  {Lee}}, \bibinfo {author} {\bibfnamefont {Y.-H.}\ \bibnamefont {Kim}},
  \bibinfo {author} {\bibfnamefont {D.~K.}\ \bibnamefont {Hwang}}, \bibinfo
  {author} {\bibfnamefont {W.~K.}\ \bibnamefont {Choi}},\ and\ \bibinfo
  {author} {\bibfnamefont {J.-Y.}\ \bibnamefont {Kim}},\ }\bibfield  {title}
  {\enquote {\bibinfo {title} {Synthesis and optoelectronic characteristics of
  20 nm diameter silver nanowires for highly transparent electrode films},}\
  }\href {https://doi.org/10.1039/C5RA25310A} {\bibfield  {journal} {\bibinfo
  {journal} {{RSC} Adv.}\ }\textbf {\bibinfo {volume} {6}},\ \bibinfo {pages}
  {11702--11710} (\bibinfo {year} {2016})}\BibitemShut {NoStop}%
\bibitem [{\citenamefont {Azani}\ and\ \citenamefont
  {Hassanpour}(2018)}]{Azani2018}%
  \BibitemOpen
  \bibfield  {author} {\bibinfo {author} {\bibfnamefont {M.-R.}\ \bibnamefont
  {Azani}}\ and\ \bibinfo {author} {\bibfnamefont {A.}~\bibnamefont
  {Hassanpour}},\ }\bibfield  {title} {\enquote {\bibinfo {title} {Silver
  nanorings: {New} generation of transparent conductive films},}\ }\href
  {https://doi.org/10.1002/chem.201804788} {\bibfield  {journal} {\bibinfo
  {journal} {Chem. - Eur. J.}\ }\textbf {\bibinfo {volume} {24}},\ \bibinfo
  {pages} {19195--19199} (\bibinfo {year} {2018})}\BibitemShut {NoStop}%
\bibitem [{\citenamefont {Azani}\ \emph {et~al.}(2019)\citenamefont {Azani},
  \citenamefont {Hassanpour}, \citenamefont {Tarasevich}, \citenamefont
  {Vodolazskaya},\ and\ \citenamefont {Eserkepov}}]{Azani2019}%
  \BibitemOpen
  \bibfield  {author} {\bibinfo {author} {\bibfnamefont {M.-R.}\ \bibnamefont
  {Azani}}, \bibinfo {author} {\bibfnamefont {A.}~\bibnamefont {Hassanpour}},
  \bibinfo {author} {\bibfnamefont {Y.~Y.}\ \bibnamefont {Tarasevich}},
  \bibinfo {author} {\bibfnamefont {I.~V.}\ \bibnamefont {Vodolazskaya}},\ and\
  \bibinfo {author} {\bibfnamefont {A.~V.}\ \bibnamefont {Eserkepov}},\
  }\bibfield  {title} {\enquote {\bibinfo {title} {Transparent electrodes with
  nanorings: A computational point of view},}\ }\href
  {https://doi.org/10.1063/1.5099933} {\bibfield  {journal} {\bibinfo
  {journal} {J. Appl. Phys.}\ }\textbf {\bibinfo {volume} {125}},\ \bibinfo
  {pages} {234903} (\bibinfo {year} {2019})}\BibitemShut {NoStop}%
\bibitem [{\citenamefont {Guennebaud}, \citenamefont {Jacob}\ \emph
  {et~al.}(2010)\citenamefont {Guennebaud}, \citenamefont {Jacob} \emph
  {et~al.}}]{eigenweb}%
  \BibitemOpen
  \bibfield  {author} {\bibinfo {author} {\bibfnamefont {G.}~\bibnamefont
  {Guennebaud}}, \bibinfo {author} {\bibfnamefont {B.}~\bibnamefont {Jacob}},
  \emph {et~al.},\ }\href@noop {} {\enquote {\bibinfo {title} {Eigen v3},}\
  }\bibinfo {howpublished} {http://eigen.tuxfamily.org} (\bibinfo {year}
  {2010})\BibitemShut {NoStop}%
\bibitem [{\citenamefont {Newman}\ and\ \citenamefont
  {Ziff}(2000)}]{Newman2000PRL}%
  \BibitemOpen
  \bibfield  {author} {\bibinfo {author} {\bibfnamefont {M.~E.~J.}\
  \bibnamefont {Newman}}\ and\ \bibinfo {author} {\bibfnamefont {R.~M.}\
  \bibnamefont {Ziff}},\ }\bibfield  {title} {\enquote {\bibinfo {title}
  {Efficient {Monte} {Carlo} algorithm and high-precision results for
  percolation},}\ }\href {https://doi.org/10.1103/PhysRevLett.85.4104}
  {\bibfield  {journal} {\bibinfo  {journal} {Phys. Rev. Lett.}\ }\textbf
  {\bibinfo {volume} {85}},\ \bibinfo {pages} {4104--4107} (\bibinfo {year}
  {2000})}\BibitemShut {NoStop}%
\bibitem [{\citenamefont {Newman}\ and\ \citenamefont
  {Ziff}(2001)}]{Newman2001PRE}%
  \BibitemOpen
  \bibfield  {author} {\bibinfo {author} {\bibfnamefont {M.~E.~J.}\
  \bibnamefont {Newman}}\ and\ \bibinfo {author} {\bibfnamefont {R.~M.}\
  \bibnamefont {Ziff}},\ }\bibfield  {title} {\enquote {\bibinfo {title} {Fast
  {Monte} {Carlo} algorithm for site or bond percolation},}\ }\href
  {https://doi.org/10.1103/PhysRevE.64.016706} {\bibfield  {journal} {\bibinfo
  {journal} {Phys. Rev. E}\ }\textbf {\bibinfo {volume} {64}},\ \bibinfo
  {pages} {016706} (\bibinfo {year} {2001})}\BibitemShut {NoStop}%
\bibitem [{\citenamefont {Akhunzhanov}, \citenamefont {Tarasevich},\ and\
  \citenamefont {Vodolazskaya}(2020)}]{Akhunzhanov2020}%
  \BibitemOpen
  \bibfield  {author} {\bibinfo {author} {\bibfnamefont {R.~K.}\ \bibnamefont
  {Akhunzhanov}}, \bibinfo {author} {\bibfnamefont {Y.~Y.}\ \bibnamefont
  {Tarasevich}},\ and\ \bibinfo {author} {\bibfnamefont {I.~V.}\ \bibnamefont
  {Vodolazskaya}},\ }\bibfield  {title} {\enquote {\bibinfo {title} {Circles of
  equal radii randomly placed on a plane: some rigorous results, asymptotic
  behavior, and application to transparent electrodes},}\ }\href
  {https://doi.org/10.1088/1742-5468/ab74cd} {\bibfield  {journal} {\bibinfo
  {journal} {J. Stat. Mech: Theory Exp.}\ }\textbf {\bibinfo {volume} {2020}},\
  \bibinfo {pages} {033202} (\bibinfo {year} {2020})}\BibitemShut {NoStop}%
\bibitem [{\citenamefont {Manning}\ \emph {et~al.}(2019)\citenamefont
  {Manning}, \citenamefont {da~Rocha}, \citenamefont {Callaghan}, \citenamefont
  {Ferreira},\ and\ \citenamefont {Boland}}]{Manning2019}%
  \BibitemOpen
  \bibfield  {author} {\bibinfo {author} {\bibfnamefont {H.~G.}\ \bibnamefont
  {Manning}}, \bibinfo {author} {\bibfnamefont {C.~G.}\ \bibnamefont
  {da~Rocha}}, \bibinfo {author} {\bibfnamefont {C.~O.}\ \bibnamefont
  {Callaghan}}, \bibinfo {author} {\bibfnamefont {M.~S.}\ \bibnamefont
  {Ferreira}},\ and\ \bibinfo {author} {\bibfnamefont {J.~J.}\ \bibnamefont
  {Boland}},\ }\bibfield  {title} {\enquote {\bibinfo {title} {The
  electro-optical performance of silver nanowire networks},}\ }\href
  {https://doi.org/10.1038/s41598-019-47777-2} {\bibfield  {journal} {\bibinfo
  {journal} {Sci. Rep.}\ }\textbf {\bibinfo {volume} {9}},\ \bibinfo {pages}
  {11550} (\bibinfo {year} {2019})}\BibitemShut {NoStop}%
\end{thebibliography}%

\end{document}